\documentclass[graybox]{svmult}

\usepackage{amsfonts}
\usepackage{epsfig}

\usepackage{mathptmx}       
\usepackage{helvet}         
\usepackage{courier}        
\usepackage{type1cm}        
\usepackage{makeidx}         
\usepackage{graphicx}        
\usepackage{multicol}        
\usepackage[bottom]{footmisc}

\makeindex             

\begin{document}

\title*{Path integral distance for data interpretation}
 
\author{D. Volchenkov}
\institute{Mathematische Physik,\at Universit\"{a}t Bielefeld, Universit\"{a}tsstra{\ss}e 25,  D-33615 Bielefeld, Germany, \email{volchenk@physik.uni-bielefeld.de}}
%
%
\maketitle

\abstract*{The process of data interpretation is always based on the implicit introduction of equivalence relations on the set of walks over the database. Every equivalence relation on the set of walks specifies a Markov chain describing the transitions of a discrete time random walk. In order to geometrize and interpret the data, we propose the new distance between data units defined  as a "Feynman path integral", in which all possible paths between any two nodes in a graph model of the data are taken into account, although some paths are more preferable than others. Such a path integral distance approach to the analysis of databases has proven its efficiency and success, especially on multivariate strongly correlated data where other methods fail to detect structural components (urban planning, historical language phylogenies, music, street fashion traits analysis, etc. ).  We believe that it would become an invaluable tool for the intelligent complexity reduction and big data interpretation.}

\abstract{The process of data interpretation is always based on the implicit introduction of equivalence relations on the set of walks over the database. Every equivalence relation on the set of walks specifies a Markov chain describing the transitions of a discrete time random walk. In order to geometrize and interpret the data, we propose the new distance between data units defined  as a "Feynman path integral", in which all possible paths between any two nodes in a graph model of the data are taken into account, although some paths are more preferable than others. Such a path integral distance approach to the analysis of databases has proven its efficiency and success, especially on multivariate strongly correlated data where other methods fail to detect structural components (urban planning, historical language phylogenies, music, street fashion traits analysis, etc. ).  We believe that it would become an invaluable tool for the intelligent complexity reduction and big data interpretation.}

\section{Introduction}

It was reporeted in a recent paper published in {\it ScienceDaily}
\cite{BigData:2013} that a full 90\% of all the data in the world has been generated over the last two years: "The volumes of data make up what has been designated '{\it Big Data}' -- where data about individuals, groups and periods of time are combined into bigger groups or longer periods of time." 
 We can surmise that 
the figures of recently collected data
 would only grow over time, 
and we shall be literally awash with data in the forthcoming years.
A wealth of available data has already led
  to a substantial paradigm shift 
in many areas of science.
In the past, 
 we usually  developed a theoretical concept 
based on strikingly simple equations 
involving only a few degrees of freedom first, 
and then waited sometimes for decades 
for the appropriate experimental data 
that can either confirm or refute the theory.
Quite often nowadays we have neither
a  reliable equation, nor any {\it a priori} concept of the 
essentially multi-dimensional,
multi time-scale, highly correlated data, being
aggregated over a system of enormous intricacy.
In most cases, an intelligent complexity reduction procedure  
has to be applied to the data first,
 so that it could then be  serve as a scaffold for the further theoretical concepts.

As the absolute majority of the Big Data
is collected and processed automatically, 
their volume makes them impossible for 
interpretation by a human operator,
 and calls for an automated interpretation technology. 
Below, we present a general framework for   
the analysis of complex data based on the concept 
of a 'Feynman path integral' distance. 
In contrast to the classical approaches, 
pondering the distance between two vertices in a graph 
as the number of edges in a shortest path connecting them
and the relation between two data units
 in a relational database through a continuous patch of local 
pair wise relations along a path over the database,
we suggest using a distance, in which {\it all} 
 possible paths between two vertices in a connected graph
 or a database are taken into account,
 although some paths are more preferable than others.
The very idea of assigning 
 an individual 'statistical weight'
to every  path 
calls for a model of random walk over the graph, 
in which a path is characterized 
by the probability to be followed by the random walkers. 
However, the random walk has even more profound relation to databases.
In Sec.~\ref{sec:data_Interpretation}, 
we discuss on that
the process of data interpretation 
is always based on the implicit introduction of some
 equivalence relations on the set of walks over the database.
The equivalence relation on the set of walks specifies
a transition operator, with respect to which all 
equivalent walks are equiprobable, that is a random walk.  
In Sec.~\ref{sec:data_Interpretation}, we also analyze 
the random walks of different scales and show 
how their properties vary 
in the structurally different environments. 
In Sec.~\ref{sec:path_int}, we use the method 
intrinsically similar to the path integral, 
for summing up all possible random walks that 
give us a key to the data geometrization. 
Further, in Sec.~\ref{sec:maze},\ref{sec:music},\ref{sec:cities},
 we provide some examples of the application 
of the path integral distance for the analysis of 
structural properties of mazes, musical compositions, and urban 
environments. In Sec.~\ref{sec:Morse}, 
 we introduce the first attaining time manifolds 
and study their topological properties (by calculating their Euler characteristics).
We conclude in Sec.~\ref{sec:discussion}.

\section{Data interpretation and scale dependent random walks \label{sec:data_Interpretation}}

The analysis of data
collected over the real world systems  
starts with the abstraction of 
independent entities
capable of representing some data aspect
distinguishable from other aspects
over the complexity of a domain.
By capturing how the entities are
allied to one another
by the binary relationships,
one converts the data into 
a relational database.
In a graph model of the relational database, 
entities and binary relationships 
can be thought of 
as {\it vertices} (nodes or states) $\mathcal{V}$ and 
{\it edges} (arcs or relationships)
 $\mathcal{E}\subseteq \mathcal{V}\times \mathcal{V},$  respectively.
We suppose that the database is finite that is 
$|\mathcal{V}|=N$ and $|\mathcal{E}|=E.$ Walks on a relational database 
can correspond
to a composite function 
acting from the source to the destination, 
  a variety of inheritance,
property - sub property, and 
ancestor-descendant
  relationships between vertices, 
 data queries (in the process of information retrieval),
the address allocation and assignment policies (in the process of data storage), a coding function on a space of genetic algorithms
into a space of chromosomes, etc. 

The process of data interpretation (or {\it classification}) 
is always based on the implicit introduction
 of certain 
{\it equivalence relations} (i.e. reflexive, symmetric and transitive)
 on the set of walks over the database  \cite{Volchenkov:2013}. 
It is worth mentioning that 
different equivalence relations 
over the sequences of identification characteristics 
lead to the different concepts of taxonomic categories.  
For example, in the famous 
 classification system developed by Swedish botanist Carl Linnaeus in the 1700s, 
living organisms have been grouped 
 into kingdoms, phyla, classes, orders, families, genera and species depending on a possessing of the certain attributes from a 
 hierarchically organized scheme,  
making organismal appearance similar and revealing their apparent relationship to other organisms.
Clearly, the decision upon which species an organism belongs to  
crucially depends upon a pragmatic choice based on the particularities of the species of concern and is by no means immutable.
Basic genetic analysis information being 
grounded on an alternative equivalence relation 
had often changed our ideas of how closely some species are related, and so 
their classification also changed accordingly.
Therefore, an interpretation does not necessary reveal a "true meaning" of the data, but rather represents a self-consistent point of view on that. 
For instance, astrology claims to explain aspects of an individual personality and predict future events in the individual life based on the premise that all same day born individuals share a similar (or equal) personality determined by the positions of the sun, moon, and other planetary objects at the time of their birth. Provided  a character of events at a given day 
 is considered as a node in the "life graph",
the "astrological" equivalence partition of life paths suggests that 
all paths of the given length $n$ starting at the same node are equivalent.

\begin{figure}[ht]
 \noindent
\begin{center}
\epsfig{file=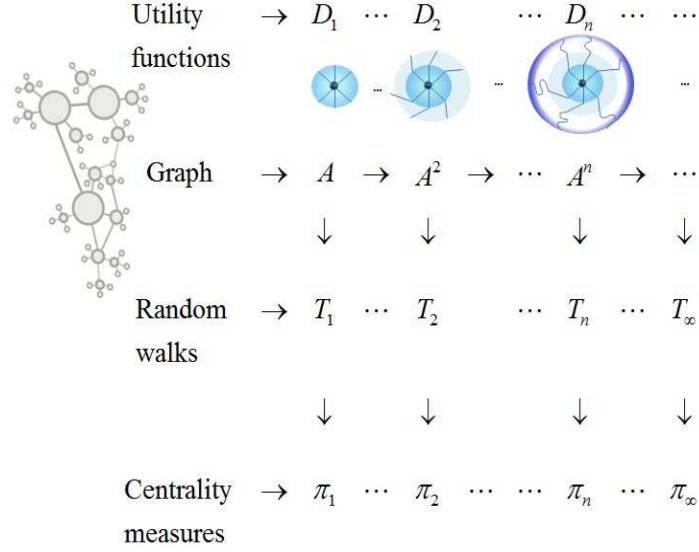, angle= 0,width =10cm, height =8cm}
  \end{center}
\caption{The "astrological" equivalence partition of paths 
 in a 
finite connected undirected  graph
(i.e., all paths of the given length $n$ starting at the same node are equivalent) induces the set of scale dependent random walks, 
the stationary distributions of which are the centrality measures accounting 
for the fractions of $n-$walks traversing a node.  }
 \label{Fig1_d}
\end{figure}

Given a probability measure 
on the set of all walks of the given length $n$, 
every equivalence relation 
over walks 
specifies 
a stochastic matrix describing 
the 
discrete time
transitions
of a 
 scale dependent {\it random walk}
between vertices $\mathcal{V}$
such that 
equivalent $n-$walks correspond to 
 {\it equiprobable} transitions \cite{Volchenkov:2013}.
For example, let us suppose that the 
 "astrological" equivalence partition of paths is applied to 
a finite connected   undirected   graph $G$ (see Fig.~\ref{Fig1_d})
described by the  adjacency matrix 
such that $A_{ij}=A_{ji}=1$ iff $i\sim j,$
 and $A_{ij}=A_{ji}=0$ otherwise.
The powers of adjacency matrix $A^n$ accounts for the numbers of paths of the length $n$ existed in the graph $G$, and the aggregated
 utility functions for each equivalence class is
given by
$D_n=\mathrm{diag}\left(\sum_{i=1}^NA^n_{1,i},\ldots,\sum_{i=1}^NA^n_{N,i}\right).$
We can define a set of stochastic transition matrices
 \begin{equation}
\label{Eq1_1}
T^{(n)}\,\,=\,\, D^{-1}_nA^n, \quad n=1,\ldots \infty
\end{equation}
that describe the discrete time 
Markov chains (the random walks) on the graph $G,$
in which all paths of length $n$
 starting at the same node are
considered equiprobable.
Provided
the adjacency matrix $A$ is irreducible,
 the Perron-Frobenius theorem
(see, for example, \cite{Graham:87})
states that
 its dominant eigenvalue 
$\alpha>0$ is simple,
 and the corresponding left and right eigenvectors
 can be chosen  positive,
$A\varphi=\alpha \varphi,$ $\varphi_i>0.$   
As $n\to\infty,$
 the elements of the transition matrices (\ref{Eq1_1})
tend to a constant row matrix, 
\begin{equation}
\label{Eq2_2}
T^{(\infty)}
\,\,=\,\,
\lim_{n\to\infty}\left(T^{(n)}\right)_{ij} 
\,\,=\,\,
\left[\frac{\varphi_i}{\sum_{i\in\mathcal{V}}\varphi_i} \right]_{i=1,\ldots,N}
\,\,=\,\,
\lim_{n\to\infty}\left( \pi^{(n)}_i\right).
\end{equation}
 The elements of the left 
 eigenvectors belonging to the biggest eigenvalue
$\mu=1$ of the matrices
  $T^{(n)}$  
are nothing else but the {\it centrality measures} 
\cite{Volchenkov:2013}
defined as the
fractions of $n$-walks traversing the node $i$, 
\begin{equation}
\label{Eq2_1}
\pi^{(n)}_i\,\,=\,\,
\frac{\sum_{j\in\mathcal{V}}A^n_{i,j}}{\sum_{s,j\in\mathcal{V}}A^n_{s,j}}.
\end{equation}
The first centrality measure,
\[
\pi^{(1)}_i\,\,=\,\, 
\frac{\sum_{j\in\mathcal{V}}A_{i,j}}{\sum_{s,j\in\mathcal{V}}A_{s,j}}\,\,=\,\,
\frac{\deg(i)}{2E},
\]
 where $\deg(i)$ is the degree of the node $i$ in $G,$
  is the well-known stationary 
distribution of the nearest-neighbor random walks; the doubled number of edges $2E$  appears in that because of each edge can be traversed in both directions.
The node centralities of the higher orders 
are increasingly  sensitive to the 
proximity of the node to irregularities and defects
of the graph.

\begin{figure}[ht!]
 \noindent
\centering
\begin{tabular}{llllll}
{\small 1.)} & \epsfig{file=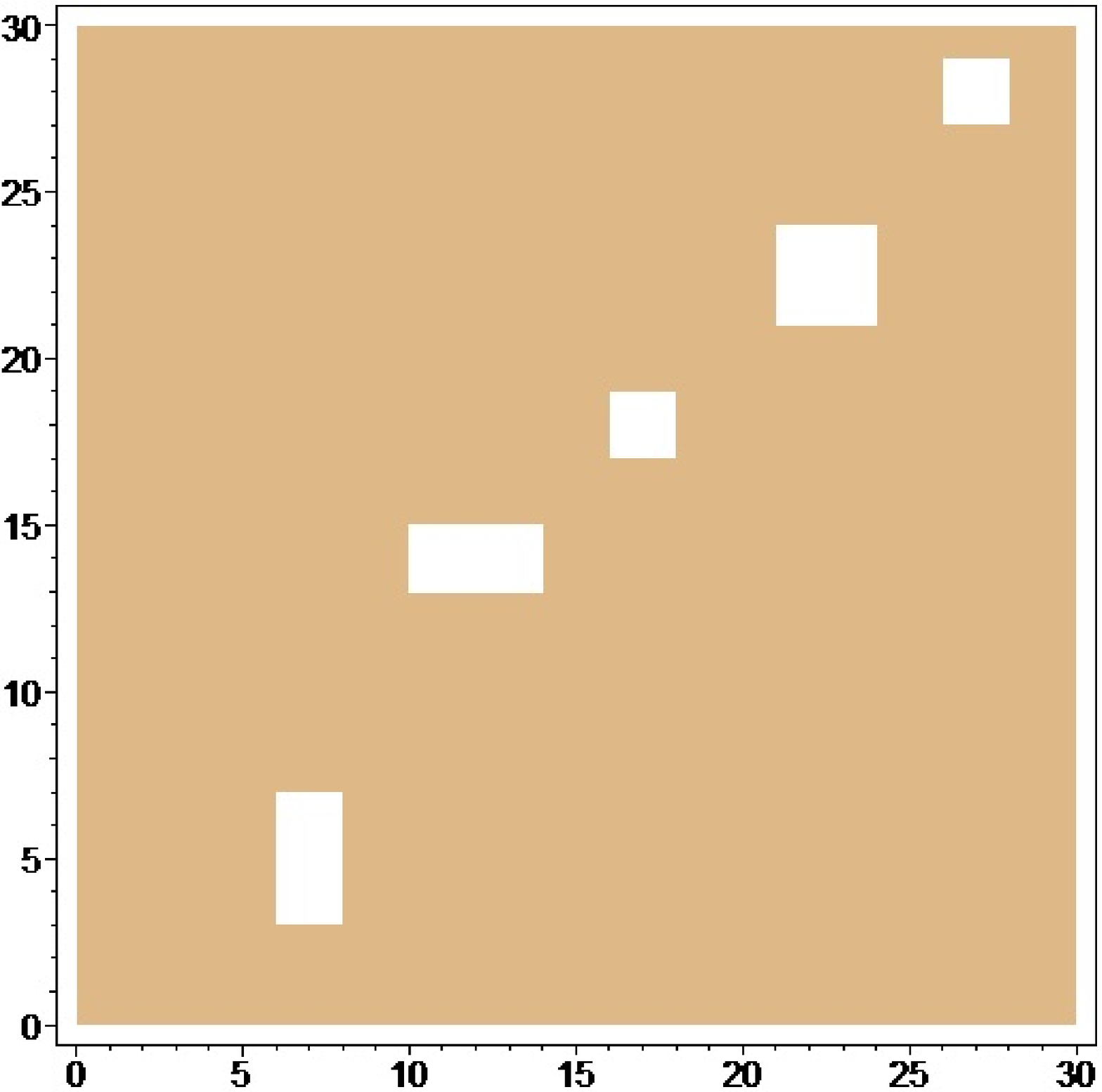, width=3.0cm, height =3.0cm} &
{\small 2.)} & \epsfig{file=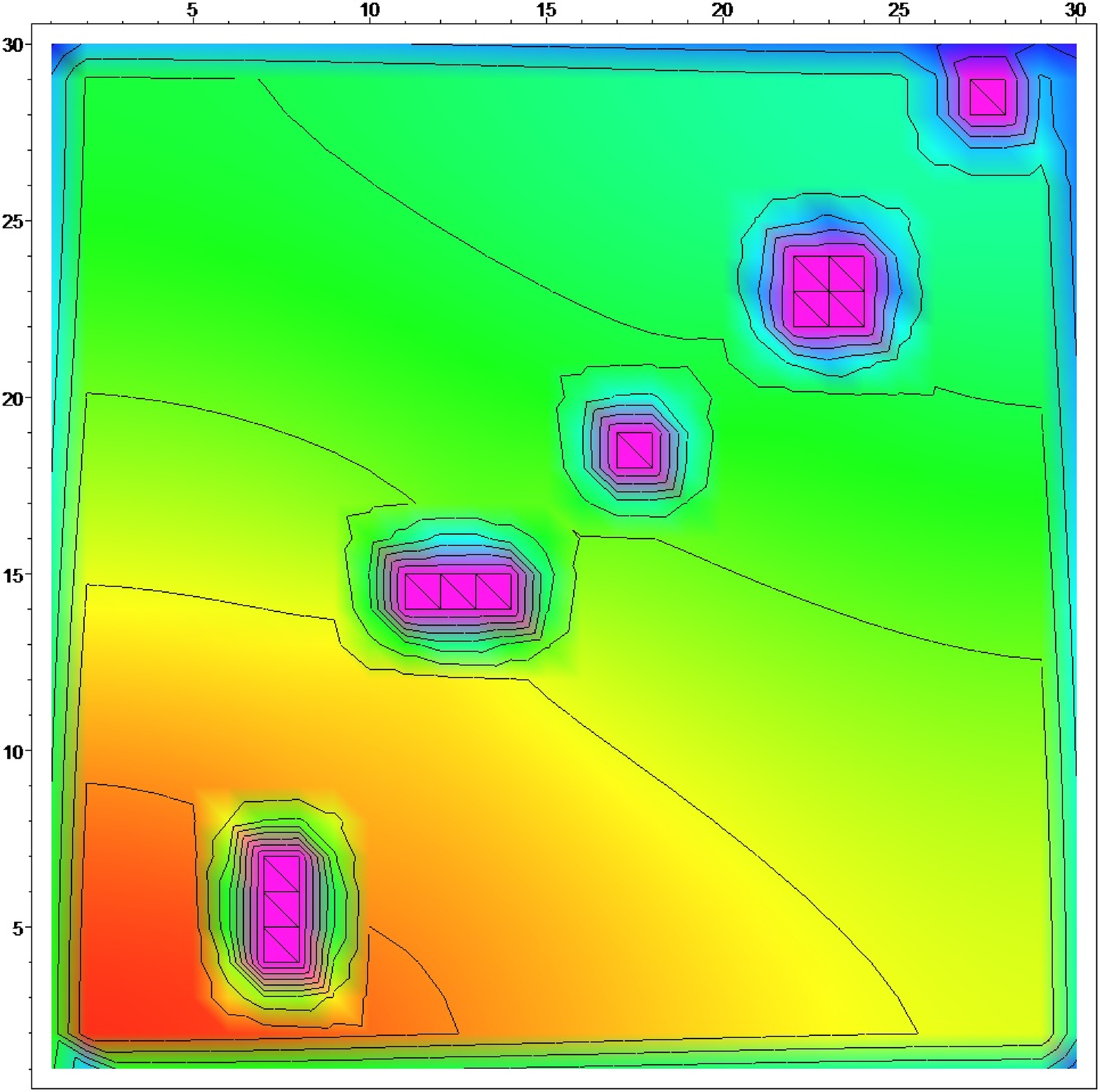, width=3.0cm, height =3.0cm} & 
{\small 3.)} &  \epsfig{file=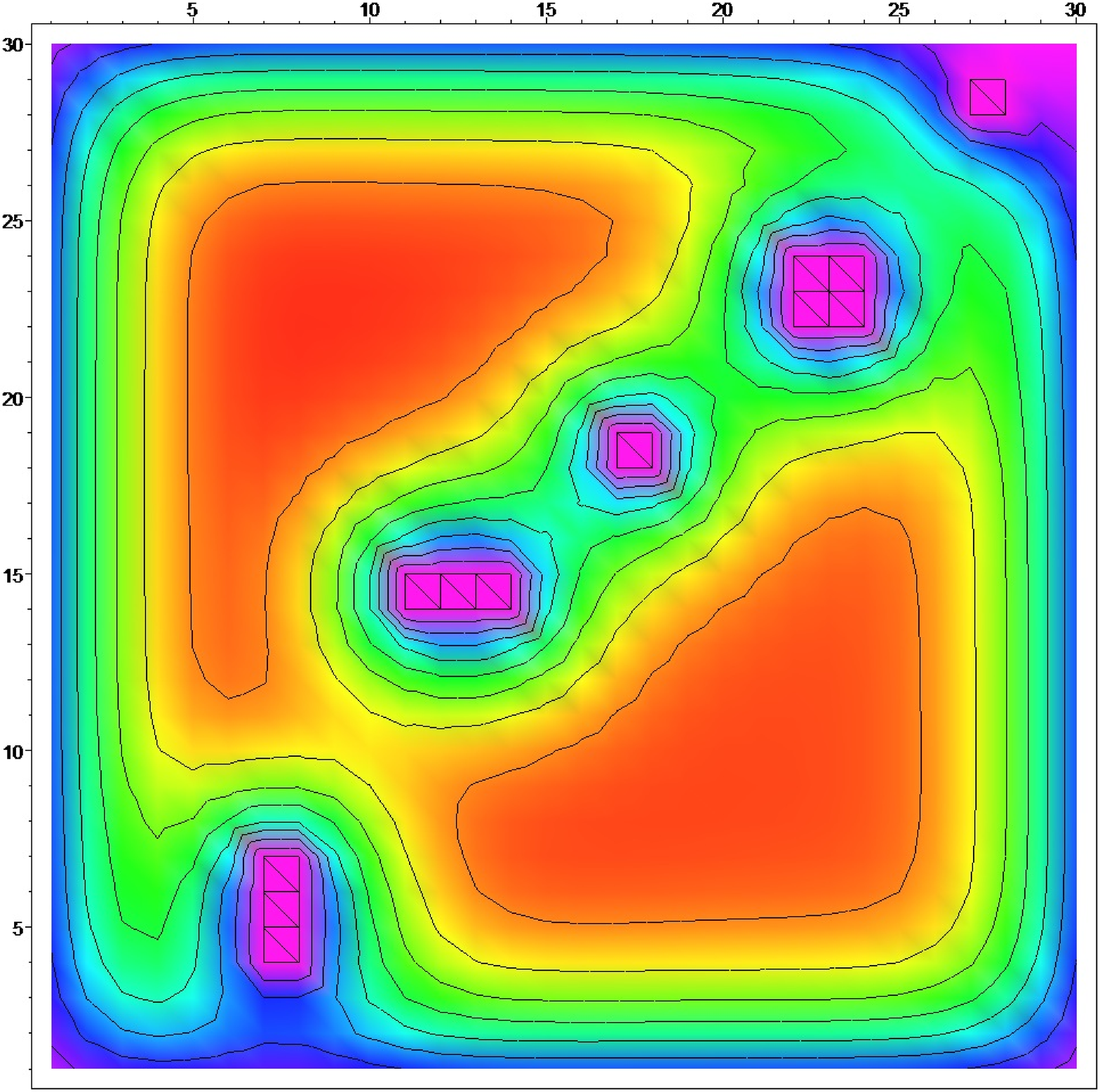, width=3.0cm, height =3.0cm} 
\end{tabular}
\caption{Two different random walks ($T^{(1)}$ and $T^{(10)}$)
 spread from the lower left corner of
1.) the square $30 \times 30$ containing a number of obstacles 
located along the main diagonal.
2.) The contour density plot representing the density of random walkers 
spreading in accordance with the usual, nearest neighbor random walk transitions $T^{(1)},$ after 100 iterations. 
3.) The contour density plot representing the density of random walkers
spreading in accordance with the  transitions to the 10-th neighbor  
$T^{(10)},$ after 100 iterations  (the stationary distribution $\pi^{(10)}$ is already reached). \label{Fig_RW}}
\end{figure}

To illustrate a dramatic dissimilarity 
of random walks at different scales  $n$, 
we show the resulting densities 
of random walkers spreading during 
100 iterations from the left lowest corner of the 
of the square $30 \times 30$ containing a number
 of rectangular obstacles located along the main diagonal (see Fig.~\ref{Fig_RW}).
 In  Fig.~\ref{Fig_RW}1, we have presented a contour density plot 
for the 
the usual nearest neighbor random walk transition operator $T^{(1)}$, 
and Fig.~\ref{Fig_RW}2 is for  $T^{(10)}$.
 
While the random walks $T^{(10)}$ have already reached the stationary distribution 
 $\pi^{(10)}$ 
by $t=100,$ the usual, nearest neighbor  random walks $T^{(1)}$ 
are still in spread toward an almost homogeneous distribution 
$\pi^{(1)} $ 
(which depend only upon the local connectivity of the cell, $\pi^{(1)}_i=\deg(i)/2E$)
being 
less sensitive either to  defects, or to the cage boundaries
and  covering all available space virtually uniformly, 
including the spaces between the obstacles.
In contrast to them, the random walks $T^{(10)}$ are 
"repelling" from obstacles, 
as the centrality measure $\pi^{(10)}$ 
quantifying the fraction of paths of the length $10$
 traversing a square cell  
 dramatically decreases 
in the vicinity of obstacles and boundaries,
 so that 
in the stationary distribution reached by $T^{(10)}$ 
the density of random walkers  between the 
obstacles and along the boundaries 
appears to be minimal. In particular, 
they almost do not diffuse beyond the last obstacle,
 at the upper right corner of the square, and between 
the first obstacle and the cage boundary, at the lower left
corner of the square  (see Fig.~\ref{Fig_RW}2).
\begin{figure}[ht!]
 \noindent
\centering
\begin{tabular}{ll}
\epsfig{file=B.eps, width=4cm, height =4cm}
& \epsfig{file=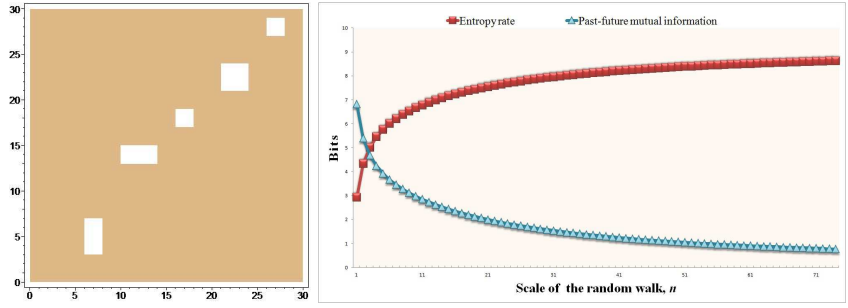, width=7cm, height =4cm} 
\end{tabular}
\caption{ Left: The square $30 \times 30$ containing a number of obstacles 
located along the main diagonal, the same we used in the previous simulation.
Right: The entropy rate (marked by squares) and the past-future mutual information 
(marked by triangles)
via the scale $n$ of the random walks defined in the square.
 \label{Fig3_RW}}
\end{figure}
Since random walks
at the different scales 
interfere with the structure 
differently, 
they can provide us with the important data
on how a structure can store, organize, and transform  
information. 
It is obvious that 
the random walks of different scales 
 $T^{(n)}$ are characterized by
the different {\it entropy rates},
\begin{equation}
\label{entropy_rate}
H^{(n)}\,\,=\,\,-\sum_{i,j\in\mathcal{V}}
\pi^{(n)}_{i}T^{(n)}_{ij}\log_2 T^{(n)}_{ij},
\end{equation}
indicating that 
the level of disorder in
the random transitions 
within the given structure
varies with the changing scale $n$.
The complimentary information about 
the level of regularities or 
structural correlations observed 
at the different scales of the structure 
is given by 
the {\it past-future mutual information}
 (the excess entropy)
introduced in \cite{Crutchfield:83,Shaw:1984},
\begin{equation}
\label{complexity_past_future}
C^{(n)}\,\,=\,\,\lim_{N\to \infty}
\left( H^{(n)}(S^N)-H^{(n)}\cdot N\right)
\,\,=\,\,-\sum_{i\in \mathcal{V}}\pi^{(n)}_{i}\,\log_2
\frac{\pi^{(n)}_i}{\,\, \prod_{j\in\mathcal{V}}\left(T^{(n)}_{ij}\right)^{T^{(n)}_{ij}}\,\,}.
\end{equation}
where the block entropy is defined by 
\begin{equation}
\label{block_entropy}
H^{(n)}(S^N)\,\,=\,\,-\sum_{S^N}P^{(n)}(S^N)\log_2 P^{(n)}(S^N),
\end{equation}
in which $P^{(n)}(S^N)$ is the probability to 
find a path (a block of symbols identifying the path) 
$S^N$ of the length $N,$ in the random walks
of the scale $n$. 
The second equality in (\ref{complexity_past_future})
 is due to the 
fact that the transition 
probability between states 
in a Markov chain 
is independent of $N$ and can be 
readily calculated ( see \cite{Li:1991} for details).

The excess entropy $C^{(n)}$ goes 
in literature by a number of 
different names (such as "complexity", \cite{Shaw:1984}) as a
measure of one type of the "memory"
that structure imposes on the process
defined on that (in our case, the random walks, 
performing transitions of the scale $n$) and thus
can serve as a measure of 
structural correlations 
 presented in the environment \cite{Feldman:2008}. 
In Fig.~\ref{Fig3_RW} (Right),  
we have presented  
the entropy rates 
calculated accordingly to \ref{entropy_rate}
and the past-future mutual information
(the  excess entropy) defined by
\ref{complexity_past_future},
 for the random walks of increasing scales 
 defined on the  square $30 \times 30$ 
including  a number of obstacles, the same as we used in the 
previous simulation (see Fig.~\ref{Fig3_RW} (Left)). 
As the scale of random walks increases, 
the level of transition randomness
quantified by the entropy rate 
$H^{(n)}$ shows the almost logarithmic growth 
 sketched on
 Fig.~\ref{Fig3_RW} (Right),
gradually approaching its maximum for 
the {\it maximum entropy} random walks
$T^{(\infty)}.$ 
At the same time, 
the behavior of the excess entropy $C^{(n)}$
is complimentary, decreasing
 from its maximal value 
attained for the nearest neighbor random walk
(the {\it maximal past-future mutual information}, or the 
{\it maximal complexity} random walks) 
 to the walks characterized by the minimal 
dependence on past path 
attained for the maximum entropy random walks 
$T^{(\infty)}.$
\begin{figure}[ht!]
  \begin{minipage}{.3\linewidth} 
\begin{flushright}
\begin{tabular}{rr}
A.) &\epsfig{file=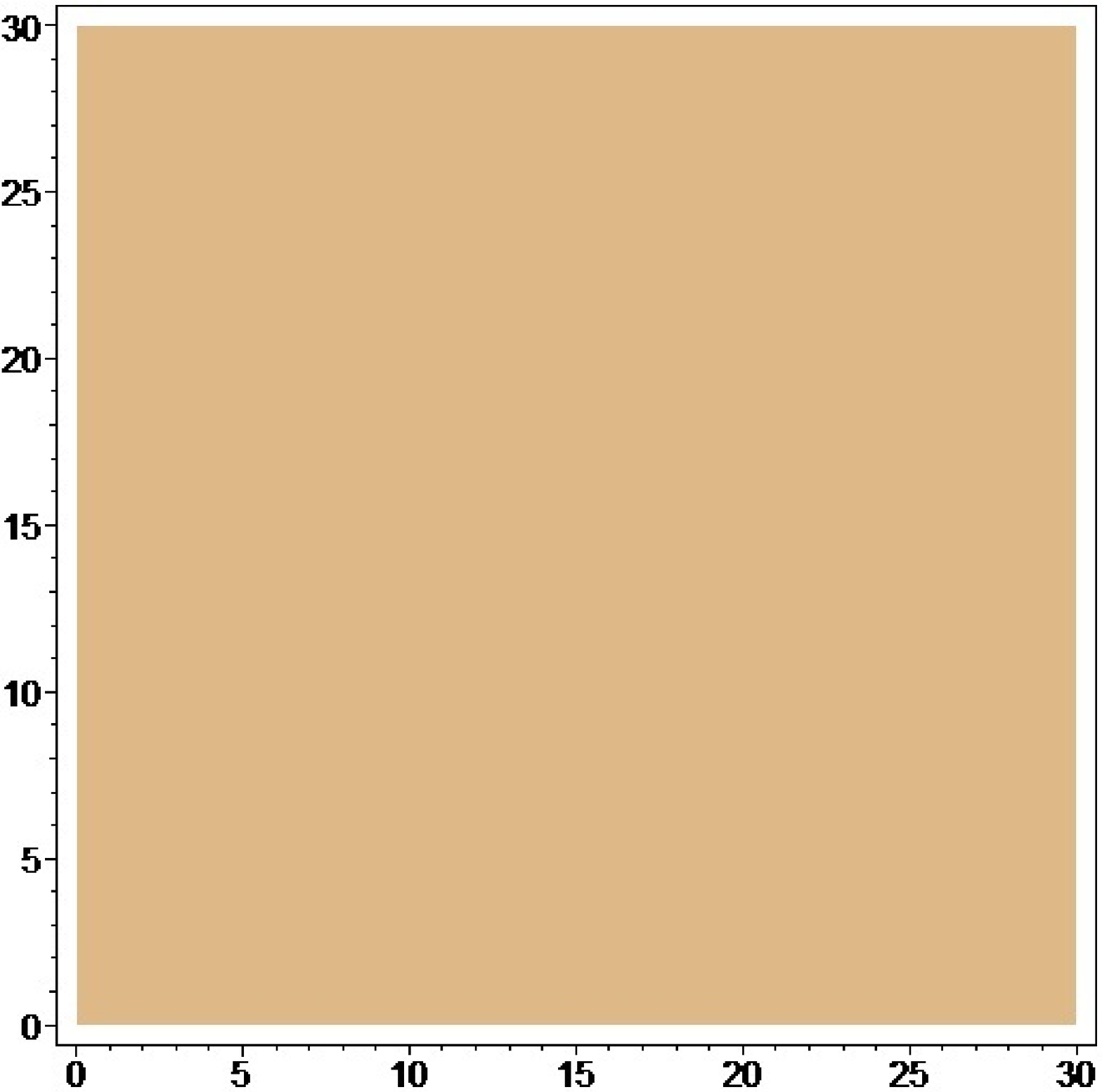, width=2cm, height =2cm} \\
B.) & \epsfig{file=B.eps, width=2cm, height =2cm} \\
C.) & \epsfig{file=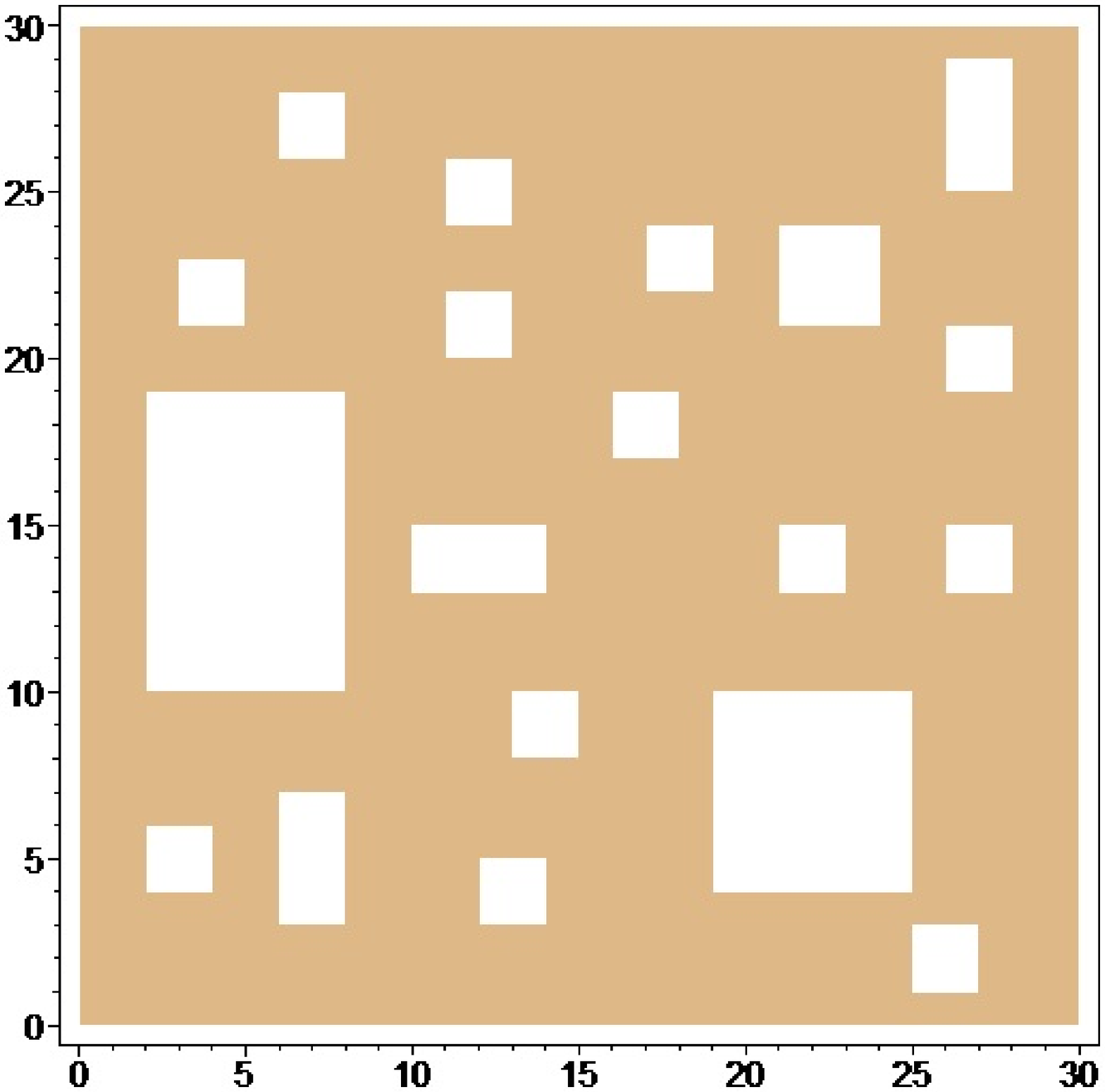, width=2cm, height =2cm} 
\end{tabular}    
\end{flushright}
\end{minipage}
\begin{minipage}{.7\linewidth}
\begin{flushleft}
\begin{tabular}{c}
\epsfig{file=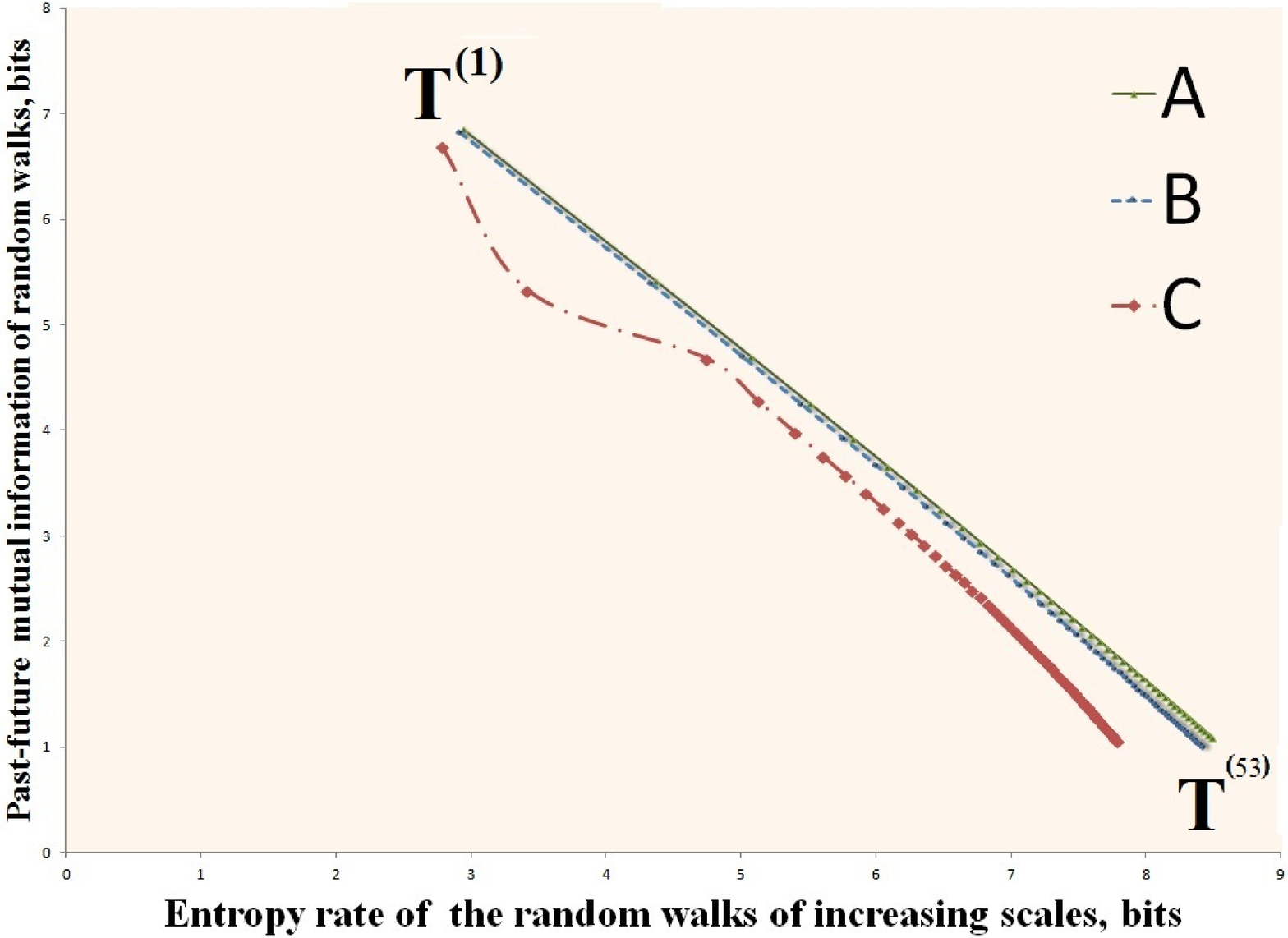, width=8cm, height =6cm}
\end{tabular}  
\end{flushleft}
\end{minipage}
\caption{ The complexity-entropy diagram shows
the past-future mutual information (vertical axis)
 versus the entropy rate 
(horizontal axis), for the  
random walks 
of increasing scales
$T^{(n)}$, for different environments:
the environment (A) constitutes the $30\times 30$ square
 free of obstacles, 
the environment (B)
contains five rectangular obstacles located along the diagonal 
of the square,
the environment (C)
 contains 20 different rectangular obstacles 
distributed randomly over the available space.
\label{Tab_music_05}}
\end{figure}
It is clear that high order (i.e., low entropy)
 does not necessarily mean a highly correlated
 structure (i.e. high past-future mutual information).
Obviously,  
the precise relation between the 
paces of the entropy rate and of
the past-future mutual information 
(complexity) of transitions for  
the random walks
across the different scales 
also depends upon the particular
 structural features 
of the environment. 
In the complexity-entropy diagram
shown in Fig.~\ref{Tab_music_05}(Right),
 we presented the 
past-future mutual information (vertical axis)
 versus the entropy rate 
(horizontal axis) for the  
random walks 
of increasing scales
$T^{(n)}.$
The complexity-entropy diagram
has been proposed by \cite{Crutchfield:83} 
and \cite{Feldman:2008}
in order to demonstrate 
how the system 
stores, organizes, and transforms information.
In our case, the diagram
Fig.~\ref{Tab_music_05} (Right)
 indicates the 
complexity-entropy relationship 
for the random walks
of  increasing scales
defined in  three different
rectangular environments 
of increasing structural complexity 
sown in Fig.~\ref{Tab_music_05} (Left).
The environment (A) is free of obstacles, 
so that the difference between the centrality measures
$\pi^{(n)}$ 
at the different scales comes merely from the boundary 
effects which do not essentially sound, 
especially for the random walks of not very large $n$.
As a consequence, the complexity-entropy relation
for the random walks defined in the 
environment (A) is almost linear, gradually 
deviating from linearity, for the large values of $n$.
Therefore we can conclude that the 
 relation between structure and order 
in the random transitions within an "empty room"
 is straightforward across the different scales: 
the structure (provided by the walls, 
bounding the environment)
 reinforces the order. 
The next environment presented in Fig.~\ref{Tab_music_05} (Left) (B)
is just the same that we used in the previous simulations; it 
contains five rectangular obstacles located along the diagonal 
of the square.
As the number of obstacles increases, 
the centralities of cells with regard to 
the long paths decay dramatically, so 
that the departure from linearity in the
 complexity-entropy relation becomes even more visible, 
especially for the large values of $n$. 
Thus, we can say that in random transitions
the strength of reinforcement of order by structure 
increases in presence of obstacles.
This statement apparently holds true 
when the number of obstacles grows, 
yet other effects can come into play.
Eventually, the last environment 
presented in Fig.~\ref{Tab_music_05} (Left) (C)
contains 20 different rectangular obstacles 
distributed randomly over the available space.
As we expected, the departure
from linearity in the
 complexity-entropy relation 
increases for the environment (C), 
and an intriguing  deviation 
 also appears for $T^{(2)},$ determining the random walks based 
on the transitions to the second nearest neighbors.
The inordinate drop of the past-future mutual information values
producing the remarkable deviation 
of the complexity-entropy relation
from linearity  (see Fig.~\ref{Tab_music_05} (Right) (C))
for the 
short-range 
random walks
 obviously signifies 
the localization of random walkers inside
 the spaces formed by the filamentary sequences of the
closely located obstacles.  
As we have already mentioned, 
the past-future mutual information 
quantifies the dependence of the forthcoming 
random transitions upon the past walk, and
therefore the decay of that value
 indicates some extend of indifference 
of random transitions regarding 
the previous history, for the short-range walks.
No matter how a random walker performing
 the short-range walk enters the 
locally quasi-bounded 
space, with the relatively high probability 
 it remains trapped in that at least for a while.  

In the next sections,
 we suppose that 
the random walk of a proper scale is chosen that 
eliminates the structural features of the environment 
or a database. We show how it is possible to
 geometrize the data with the use of the random walks.

\section{Path integral distance and probabilistic geometry \label{sec:path_int}}
\noindent

The path integral formulation 
of quantum mechanics 
has been proposed by Richard Feynman
as a description of quantum theory which generalizes the action principle of classical mechanic: 
the classical notion of a single, unique trajectory for a system 
is replaced in that with a sum, or functional integral, 
over an infinity of possible trajectories to compute a quantum amplitude
called the {\it propagator}.
The path integral also relates quantum and stochastic processes,
as 
the Schr\"{o}dinger equation is nothing else but 
a diffusion equation with an imaginary diffusion constant,
and therefore 
the path integral is an analytic continuation of a method for summing up all possible random walks.
The propagator 
resulted from the path integral calculation 
 also can be viewed as the left inverse of the operator
 appropriate to the spread of particles,
 and is therefore considered to be 
the {\it Green's function} of the Laplace operator
describing the diffusion process. 
It is well known that if the kernel of the diffusion operator is non-trivial,
 then the Green's function is not unique. 
However, in practice, 
the certain symmetry requirement 
imposed on the kernel will give a unique Green's function. 
It is known that 
Green's functions provide a powerful tool in dealing with a wide
range of  diffusion-type
problems such as 
chip-firing games, load balancing algorithms,
and 
the calculation of the so-called {\it hitting time}, 
the expected number of steps for 
a Markov chain to reach a state $y$ with an initial state $x$, \cite{Chung:2000}. 
Below, we briefly describe the Green's function 
approach to  random walks defined on graphs or databases,
following \cite{Volchenkov:2011,Volchenkov:2013}.

Given a random walk defined by a transition matrix $T$
on a finite connected undirected weighted graph 
$G(\mathcal{V},\mathcal{E})$, all vertices 
and their subsets can be 
characterized by 
certain probability distributions 
and characteristic times 
\cite{Volchenkov:2011}.
The stationary distribution 
of random walks 
(the left eigenvector of the transition matrix 
$ T$ belonging to the maximal eigenvalue $\mu=1$)
determines a unique measure on $\mathcal{V}$
with respect to which the 
 transition operator 
$T$
becomes self-adjoint and is
represented by a symmetric transition matrix $\widehat{T}.$
 The use of  self-adjoint operators (such as the normalized graph Laplacian)
becomes now standard in spectral graph theory
\cite{Chung:1997} and in studies 
devoted to random walks on graphs
 \cite{Lovasz:1993}.
Diagonalizing the symmetric matrix $\widehat{T}$,
we obtain 
$
\widehat{T}={\Psi} { M}{\Psi}^\top,
$
where ${\Psi}$ is an orthonormal matrix,
$
{\Psi}^{\top}={\Psi}^{-1},
$
 and $ M $ is a
 diagonal matrix with entries 
$1=\mu_1>\mu_2\geq\ldots\geq\mu_N> -1$
(here,
 we do not consider bipartite graphs,
 for which $\mu_N=-1$).
The rows 
${\psi}_k=\left\{
\psi_{k,1},\ldots,\psi_{k,N}
\right\}$
 of the orthonormal matrix
${\Psi}$
are the real
eigenvectors of 
$\widehat{T}$
that forms
an orthonormal
 basis
in Hilbert space $\mathcal{H}(\mathcal{V}),$
${\psi}_k:\mathcal{V}\to S_1^{N-1},$ $k=1,\ldots N$,
where $S_1^{N-1}$ is the $N-1$-dimensional unit sphere.
We consider 
the eigenvectors ${\psi}_k$ ordered 
in accordance to the  eigenvalues they belong to.
For eigenvalues of  algebraic multiplicity $m>1$,
 a number of
linearly independent 
orthonormal
 ordered
 eigenvectors can be chosen to span
the associated eigenspace.
The first eigenvector
    $ {\psi}_1$ belonging to
 the largest eigenvalue $\mu_1=1$
(which is simple)
is the Perron-Frobenius eigenvector 
that determines the
stationary distribution of random walks
over the graph nodes,
$ \psi_{1,i}^2=\pi_i.$

The diffusion process is 
 described by the irreducible  
Laplace operator 
${L}={1} - { T}$
which has the one-dimensional null space
spanned by the vector  $\pi.$ 
As being a member of 
the multiplicative group
under the ordinary matrix multiplication
 \cite{Erdelyi:1967,Meyer:1975}, 
the Laplace operator 
possesses a {\it group inverse} 
 (a special case of
 {\it Drazin inverse},
 \cite{Drazin:1958,Ben-Israel:2003,Meyer:1975})
with respect to this group, 
$L^\sharp,$
 which 
satisfies the following conditions \cite{Erdelyi:1967}:
$$
 {  LL}^\sharp {  L} = {  L}, \quad
   {  L}^\sharp {  LL}^\sharp = {  L}^\sharp, 
\quad \mathrm{and}\quad
 \left[{  L},{  L}^\sharp\right] = 0,
$$ 
where $[{  A},{  B}]={  AB}-{ BA}$ denotes
 the commutator of 
  matrices.
The last condition
implies that 
${  L}^\sharp$
shares the same set of symmetries 
as the Laplace operator, 
being the Green function of the diffusion equation.
 The methods for computing the group generalized inverse 
for matrices of $\mathrm{rank}({ L})=N-1$
have been developed in \cite{Robert:1968,Campbell:1976}
and by many other authors. 
Perhaps, the most 
elegant way is by considering the 
eigenprojection of the matrix $ L$
corresponding to the eigenvalue $\lambda_1=1-\mu_1=0$
developed in \cite{Campbell:1976,Hartwig:1976,Agaev:2002}, 
\begin{equation}
\label{Group_inverse_Agaev}
{  L}^\sharp\,\,=\,\,\left({  L}+{  Z}\right)^{-1}-{  Z},
\quad { Z}\,\, =\,\,\prod_{\lambda_i\ne 0}
\left({  1}-L/{\lambda_i}\right),
\quad \lambda_i\,\,=\,\,1-\mu_i
\end{equation}
where the product in the idempotent matrix ${ Z}$
is taken over all nonzero eigenvalues $\lambda_{i>1}$ of $  L.$
The eigenprojection 
(\ref{Group_inverse_Agaev}) can be considered as 
a 
stereographic projection  
that projects the points 
${\psi}_k$ on the 
 sphere $S_1^{N-1}$ 
to a projective manifold
such that
all vectors 
collinear to the 
vector  $\psi_1$
(corresponding to the 
stationary distribution
of random walks)
 are projected onto a common image point.
Since 
$\psi_{1,i}\equiv\sqrt{\pi_i}>0$ for any
$i\in \mathcal{V},$
we 
can define the
new basis vectors
$\left\{\psi'_{k}\right\}_{k=1}^N,$
spanning  
the projection space $\mathrm{P}\mathbb{R}_{\pi}^{(N-1)},$
such that 
$\psi'_{k}=\left(1,{\psi_{2,i}}/{\psi_{1,i}},\ldots,
{\psi_{N,i}}/{\psi_{1,i}}\right).$
We define the inner product 
 between
 any two vectors  
$\xi,\zeta \in \mathbb{R}^{N}$
by
\begin{equation}
\label{inner_PRODUCT}
\left(\xi,\zeta\right)_{T}\,\, =\,\,
 \left(\xi,{  L}^\sharp\zeta\right).
\end{equation}
 The inner product (\ref{inner_PRODUCT})
is a symmetric real valued scalar function that
allows us to define the (squared)
norm of a vector $\xi$  by
\begin{equation}
\label{norm}
\left\|\, \xi\,\right\|^2_{T}\,\, =\,\, \left(\xi,{  L}^\sharp\xi\right)
\end{equation}
and an angle 
 $\theta\in [0,180^{\mathrm{o}}]$
 between two vectors,
\begin{equation}
\label{angle}
\theta \,\,=\,\, \arccos\left( \frac{ \left(\xi,\zeta\right)_{T} }{ \left\|\, \xi\,\right\|_{T}\left\|\, \zeta\,\right\|_{T}}\right).
\end{equation}
 The Euclidean distance between two vectors
is given by
\begin{equation}
\label{Euclidean_distance}
\left\|\xi-\zeta\right\|^2_{T}\,\, =\,\, \left\|\, \xi\,\right\|^2_{T}+\left\|\, \zeta\,\right\|^2_{T}-2\left(\xi,\zeta\right)_{T}.
\end{equation}
 For instance, 
let us consider the vector (distribution)
 $\mathbf{e}_i=\{0,\ldots 1_i,\ldots 0\}$ 
pointing at the vertex $i$ of the graph $G$
in the canonical basis. The 
spectral representation of the generalized 
inverse for undirected graphs \cite{Volchenkov:2011} is given by 
\begin{equation}
\label{L_sharp}
{  L}^\sharp_{i,j}\,\,=\,\,
\sum_{k=2}^N \frac 1{\lambda_k}\frac{\psi_{i,k}}{\psi_{i,1}}\frac{\psi_{j,k}}{\psi_{j,1}},\quad
i,j=1,\ldots,N,
\end{equation}
in which $\lambda_k$ are all nontrivial eigenvalues of the
Laplace operator.
The matrix (\ref{L_sharp}) is real symmetric 
semi-positive, as its smallest eigenvalue $\Lambda_1=0.$
Accordingly (\ref{norm}), the 
spectral representation for
 the squared norm of  $\mathrm{e}_i$
equals
\begin{equation}
\label{norm_node}
f_i\,\,=\,\,\left\|\,\mathrm{e}_i\,\right\|_T^2\, =\,\frac 1{\pi_i}\,\sum_{s=2}^N\,
\frac{\,\psi^2_{s,i}\,}{\,\lambda_s\,}.
\end{equation}
 In the theory of random walks on undirected 
graphs
\cite{Lovasz:1993},
the latter result is known
 as the spectral representation 
of the   {\it first passage time}
to the node $i\in \mathcal{V}$,
the expected number of steps required to
reach the node $i $ for the first time 
(i.e. without visiting any node twice)
starting from a node randomly chosen
among all nodes of the graph
accordingly to the stationary distribution $\pi$.
It is important to mention that 
the norm (\ref{norm}) 
of the stationary distribution $\pi$ equals zero,
\begin{equation}
\begin{array}{ll}
f_\pi\,\,=\,\,
\left\|\psi_1^2\right\|_T^2 & =
\sum_{i=1}^N\psi_{1,i}^2
\sum_{1,j}^N\psi_{1,j}^2
\sum_{\alpha=2}^N\frac 1{\lambda_\alpha}\frac{\psi_{\alpha,i}}{\psi_{1,i}}\frac{\psi_{\alpha,j}}{\psi_{1,j}} \\
& =
\sum_{\alpha=2}^N \frac 1{\lambda_\alpha}
\underbrace{\sum_{j=1}^N\psi_{1,j} \psi_{\alpha,j}}_0
\underbrace{\sum_{i=1}^N\psi_{1,i} \psi_{\alpha,i}}_0
=0,
\end{array}
\label{trivial_norm_stationary_distr}
\end{equation}
due to 
the orthogonality of eigenvectors $\psi_k$.

The Euclidean distance between any two nodes
of the graph induced by the random walk,    
\begin{equation}
\label{commute}
K_{i,j}\,=\,\left\|\,\mathrm{e}_i-\mathrm{e}_j\,\right\|^2_T\,=\, \sum_{s=2}^N\,
\frac 1{\lambda_s}\left(\frac{\psi_{s,i}}{\sqrt{\pi_i}}-\frac{\psi_{s,j}}{\sqrt{\pi_j}}\right)^2,
\end{equation}
is nothing else but the {\it commute time},
 the expected number of
steps required for a random walker starting at $i$ to
visit $j$ and then to return back to $i$,
without visiting any node twice
\cite{Lovasz:1993}.
The commute time can be represented as a sum,
$K_{i,j}=H_{i,j}+H_{j,i}$, in which
\begin{equation}
\label{hitting}
H_{i,j}\,\,=\,\,\left\|\,\mathrm{e}_i\,\right\|^2_T - \left(\,\mathrm{e}_i,\mathrm{e}_j\,\right)_T
\end{equation}
is the   {\it first-hitting time} which is the expected number
of steps a random walker starting from the node $i$ needs to reach
$j$ for the first time,  \cite{Lovasz:1993}.
The first-hitting time satisfies the equation
\begin{equation}
\label{access_eq}
H_{ij}\,=\,1+\sum_{i\sim v}H_{vj}T_{vi}
\end{equation}
reflecting the fact that
the first step takes a
 random walker to
a neighbor $v\,\in\,V$ of the starting node $i\,\in\, V$,
and then it has to reach the node $j$ from there.
In principle,
the latter equation
 can be directly used for
computing of the first-hitting times,
however, $H_{ij}$ are not
 the unique solutions of
(\ref{access_eq});
 the
correct definition requires
an appropriate diagonal boundary
condition, $H_{ii}=0$,
for all $i\,\in \,V$, \cite{Lovasz:1993}.

\begin{figure}[ht]
 \noindent
 \begin{center}
 \epsfig{file=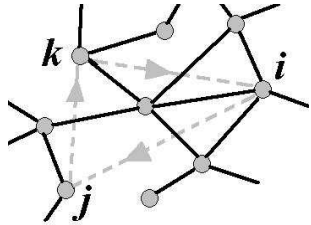,  angle= 0,width =4cm, height =3cm}
 \end{center}
\caption{ The triangle symmetry of the first-hitting times:
the sum of first-hitting times calculated for  random walks
defined by the
transition operator of  nearest-neighbor random walks visiting any three nodes $i$, $j$, and $k$,
equals to the sum of the first-hitting times
 in the reversing direction.}
\label{Fig2_04}
\end{figure}

The zero-diagonal matrix of first-hitting times
is not symmetric,
 $H_{i,j}\ne H_{j,i}$,
 even  for a regular graph.
However, a deeper  {\it triangle symmetry property}
 (see
Fig.~\ref{Fig2_04}) has been observed in \cite{Coopersmith:1993}
for random walks defined by the transition operator
 of   nearest-neighbor random walks.
Namely, for every three nodes in the graph, the consecutive sums of
the first-hitting times in the clockwise and in the
counterclockwise directions are equal,
\begin{equation}
\label{triangle}
H_{ij}+H_{jk}+H_{ki}\,=\,H_{ik}+H_{kj}+H_{ji}.
\end{equation}
We can now use the first-hitting times
in order to quantify the accessibility of nodes
and subgraphs for random walkers.
Its average
 with respect to
the first index
equals
the first-passage time to the node,
\begin{equation}
\label{mean_heating}
f_j\,\,=\,\,
\left\| \mathrm{e}_j \right\|_T^2 \,\, = \,\,
\sum_{i \in   \mathcal{V}} 
\pi_i H_{i,j}.
\end{equation}
It is worth a mention that 
matrix (\ref{L_sharp}) can be considered 
as the Gram matrix,
$L^{\sharp}=\left({\mathbf{q}}_i,{\mathbf{q}}_j\right)_{P\mathbb{R}^{(N-1)}},$
with respect to the usual dot product 
 of vectors 
in the projection space $P\mathbb{R}^{(N-1)}.$
The vector 
\begin{equation}
\label{dot_product}
{\mathbf{q}}_i\,\,=\,\,\left\{
\frac 1{\sqrt{\lambda_2}}\frac{\psi_{2,i}}{\psi_{1,i}},\ldots,
\frac 1{\sqrt{\lambda_N}}\frac{\psi_{N,i}}{\psi_{1,i}}
\right\}
\end{equation}
represents an image of the vertex $i\in \mathcal{V}$
in the projection space $P\mathbb{R}^{(N-1)}.$
The image of the graph $G(\mathcal{V},\mathcal{E})$
 in the projection space $P\mathbb{R}^{(N-1)}$ 
constitutes a diffusion manifold
of self-avoiding random walks 
(in which nodes cannot be visited twice)
 in affine subspace, 
as we can subtract vertices
(by component wise subtracting of their images (\ref{dot_product}))
 to get vectors, or add a vector to a vertex to get another vertex,
 but we cannot add new vertices. 
It seems natural to 
 describe the structural properties of the graph 
using the topology of the manifold
of self-avoiding diffusion 
in the projection space $P\mathbb{R}^{(N-1)}.$
The scalar product $\left(\mathbf{e}_i,\mathbf{e}_j\right)_{P\mathbb{R}^{(N-1)}}$ 
estimates the
expected overlap of random paths toward the nodes
 $i$ and $j$ starting from a
node randomly chosen in accordance with 
the stationary distribution of random walks $\pi$ \cite{Volchenkov:2011}.
The normalized expected overlap of random paths
 given by the cosine of an angle (\ref{angle}) calculated in
the $(N-1)-$dimensional Euclidean space associated to random walks
 has the structure of
Pearson's coefficient of linear correlations
 that reveals it's natural
statistical interpretation.
If the cosine of  (\ref{angle}) is close to 1,
the expected random paths toward the 
both nodes are mostly identical.
The value of cosine is close to -1 if the walkers
share the same random paths but in the opposite direction.
Finally, the  correlation coefficient equals 0
if the expected random paths toward the nodes do not overlap
at all.

\section{Maze of labyrinths \label{sec:maze}}

\begin{figure}[ht!]
 \noindent
\begin{center}
\begin{tabular}{lr}
\epsfig{file=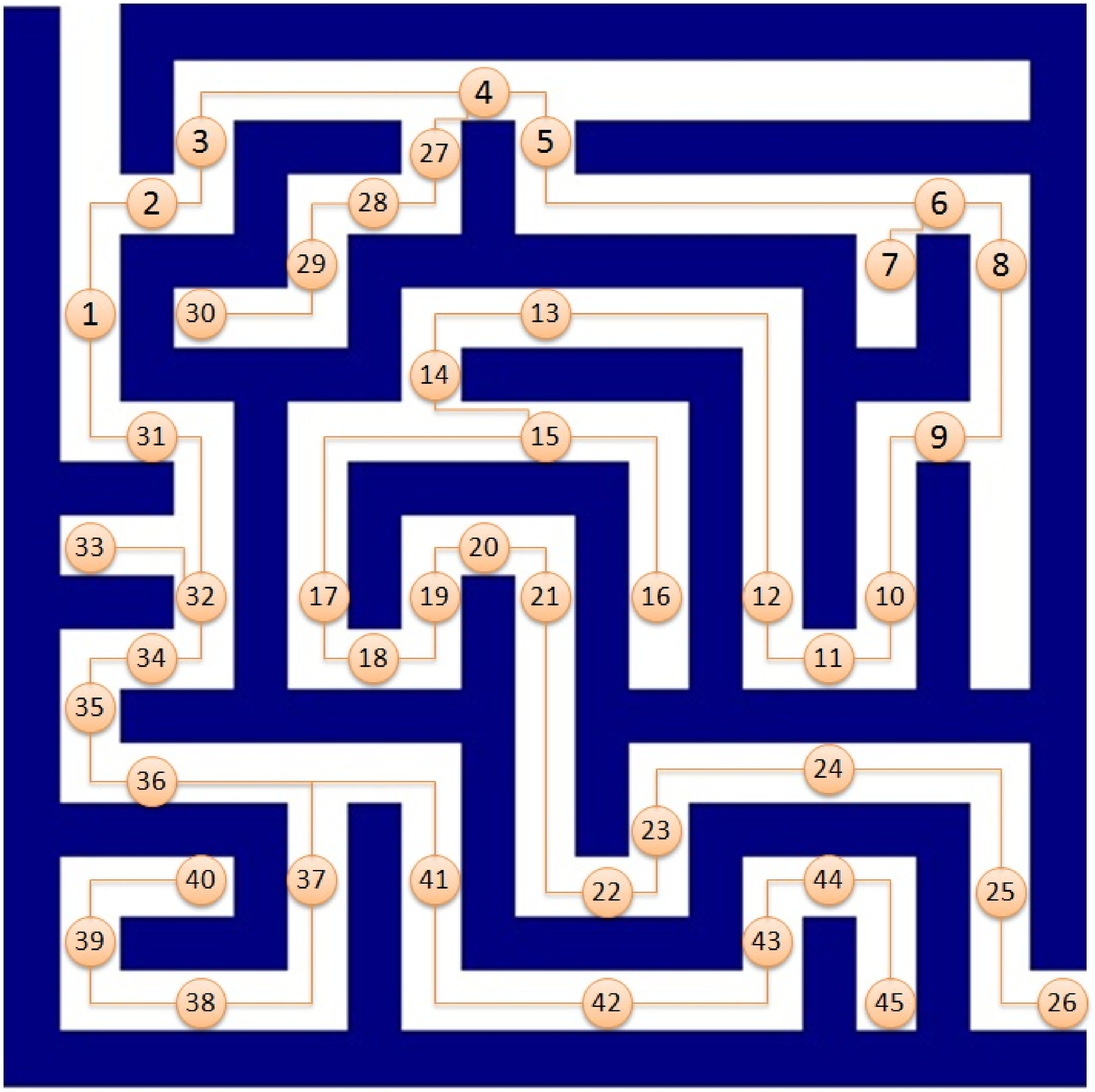, width=5cm, height =5cm}& 
 \epsfig{file=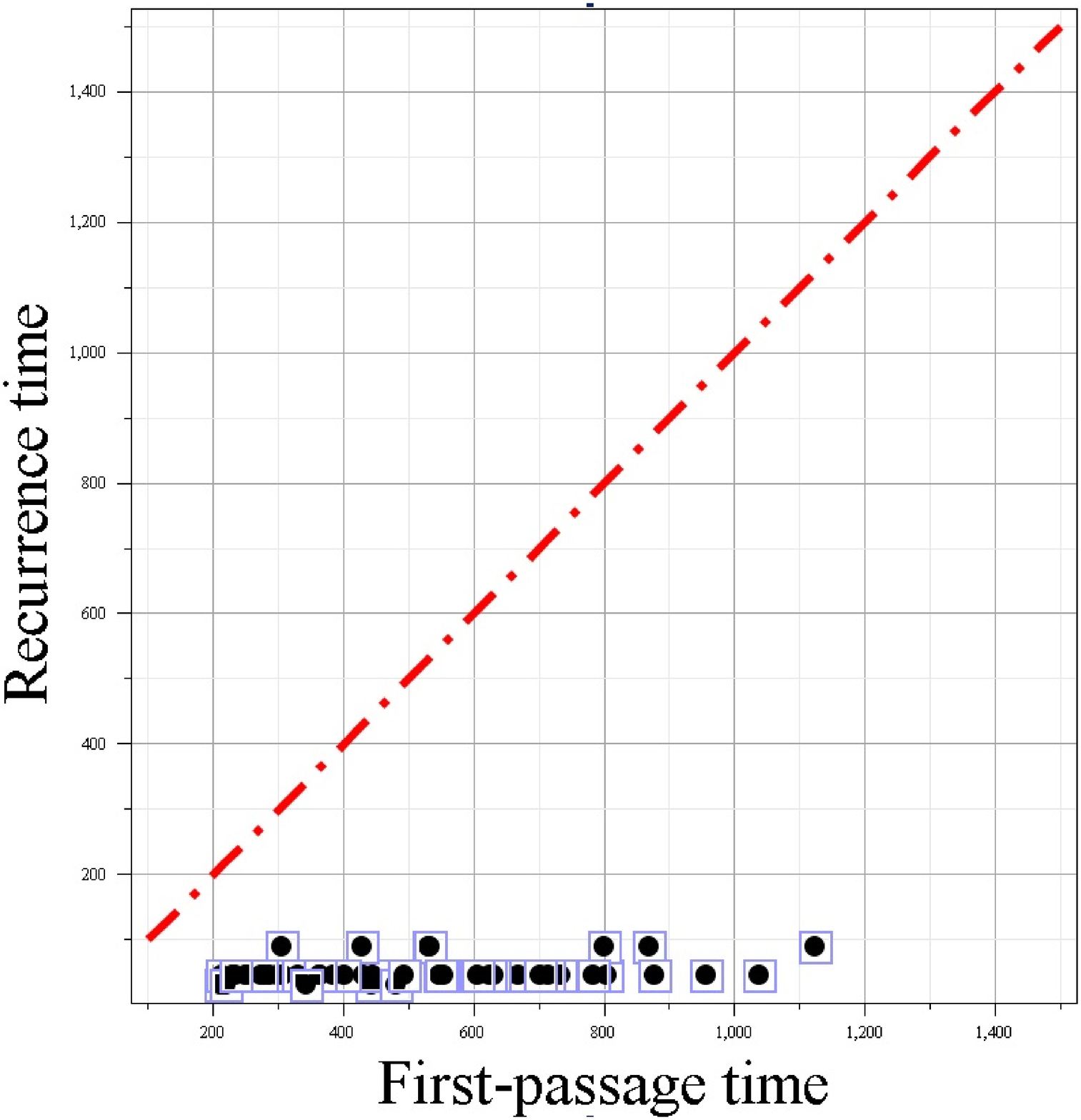, width=5.5cm, height =5.5cm} 
\end{tabular}
\end{center}
\caption{ Left: the maze consisting of 
45 interconnected rectangles
providing the space for motion,
along its spatial graph 
representing the connectivity pattern.
Right: the recurrence times to all 45 rectangular spaces of the maze 
via the first-passage times to them (both characteristic 
times are calculated with regard to the nearest-neighbor random walks). The dash-dotted reference line 
($x=y$) indicates 
 the equality.
 \label{Fig_maze}}
\end{figure}

It is believed that once upon a time labyrinths had served as traps for malevolent spirits or as defined paths for ritual dances 
(there are surviving descriptions of French clerics performing a ritual Easter dance along the path on Easter Sunday \cite{Kern:2000}). 
 The present-day notion of a labyrinth is a place
 where one can lose his way,  a confusing path, hard to follow without 
a thread, a intricate and inextricable path to the home of a sacred ancestor
 \cite{Schuster:1996}. 
Being a symbol of
ambiguity and disorientation, the notion of  
a labyrinth is also used to describe a confusing logic of arguments.
In Plato's dialogue {\it Euthydemus}, Socrates describes the labyrinth in
the line of a logical argument:
\begin{quote} 
"Then it seemed like falling into a labyrinth: we thought we were at the finish, but our way bent round and we found ourselves as it were back at the beginning, and just as far from that which we were seeking at first."
\end{quote}
It is interesting to discuss the general structural properties 
of labyrinths that make them so difficult to navigate in and, at 
the same time, so mysteriously attractive to our minds. 
In Fig.~\ref{Fig_maze} (left), we have presented 
the maze consisting of 
45 interconnected rectangles
providing the space for motion
 and displayed  its spatial graph 
representing the connectivity pattern.

It is known that  for a stationary,
discrete-valued stochastic process
 the expected recurrence time to return to a state
is the reciprocal of
 the probability of this state, \cite{Kac:1947}.
The expected {\it recurrence time}
to a node
which 
indicates
how long a random walker must wait
to revisit the site 
 is inverse proportional to the stationary distribution,
$\pi_i$,
\begin{equation}
\label{Alice_recurrence}
r_i 
   \,\,=\,\, \frac 1{\,\pi_i\,} \,\,=\,\,\frac 1{\,\psi^2_{1,i}\,} .
\end{equation}
It follows from the spectral representation 
of the first-passage time (\ref{norm_node}) that 
they are proportional to each other,
\begin{equation}
\label{recurrence_fpt_proportionality}
f_i\,\,=\,\,\left\|{\bf e}_i\right\|^2_T\,\,=\,\,
\frac{1}{\psi^2_{1,i}}
\sum_{k=2}^N\frac{\psi^2_{k,i}}{\lambda_k}\,\,=\,\,
r_i\cdot\sum_{k=2}^N\frac{\psi^2_{k,i}}{\lambda_k},
\end{equation}
where the proportionality coefficient can broadly vary. 
In particular,
for the nearest-neighbor random walks,
 it can be seen from the Fig.~\ref{Fig_maze}(Right)
 that the first-passage times to places in mazes 
are systematically {\it longer} than 
the recurrence times to them. 
Although it may take a quite long time for a random walker
to reach a place for the first time,
  the walker is doomed to revisit this place 
again and again, as the recurrence time to that
 might be quite short, as being 
determined (for the 
the nearest-neighbor random walks)
merely by the 
{\it local} connectivity of the place
 (that might be quite high), 
but not by the integral features 
relating the place to the global structure 
of the environment. 
In fact, the random walker appears to be trapped 
 within the maze environment and might find it confusing. 
In Fig.~\ref{Fig_maze}(Right), we can see that with respect to 
the overall structure of the maze, all spaces of movement 
(rectangles) constitute such the traps 
(as the first-passage times to all of them 
substantially exceed the recurrence times). 
It is clear that 
movements of a real
 human traveler
are rather self-determined than
random. However, 
as we have discussed in Sec.~\ref{sec:data_Interpretation}, 
the {\it interpretation} of a place with respect to 
the overall structure of the environment 
can be 
 based on some random walks defined in that. 
In particular, the nearest-neighbor random walks are
found practically efficient for the prediction of human navigation 
in urban environnements \cite{Blanchard:2009} 
(see also Sec.~\ref{sec:cities}). 
In the forthcoming sections, we show that 
 random walks can certainly inform us about
 the intelligibility of environments (lucidly indicated 
 by the relative land prices) 
and tonality of music melodies.

\section{Can we hear first-passage times? \label{sec:music}}
\noindent

A system for using dice 
to compose music randomly, 
without having to know neither the techniques of composition,
nor the rules of harmony, 
named {\it Musikalisches W\"{u}rfelspiel} 
- {\it Musical dice game} (MDG)
had become quite popular 
throughout Western Europe 
in the $18^{th}$ century \cite{Noguchi:1996}.
Depending upon the results of dice throws, 
the certain pre-composed bars of music 
were patched together 
resulting in different, 
but similar, 
musical pieces. 
 "{\it The Ever Ready Composer of Polonaises and Minuets}" 
was devised
by Ph. Kirnberger,
 as early as in 1757. 
The famous chance music machine 
 attributed to W.A. Mozart (\verb"K 516f")  
consisted of numerous two-bar fragments of music 
named after the different letters of the Latin alphabet
and destined to be combined together 
either at random, 
or following an anagram of your beloved
had been known since 1787.

\begin{figure}[ht!]
 \noindent
\begin{center}
\begin{tabular}{lr}
\epsfig{file=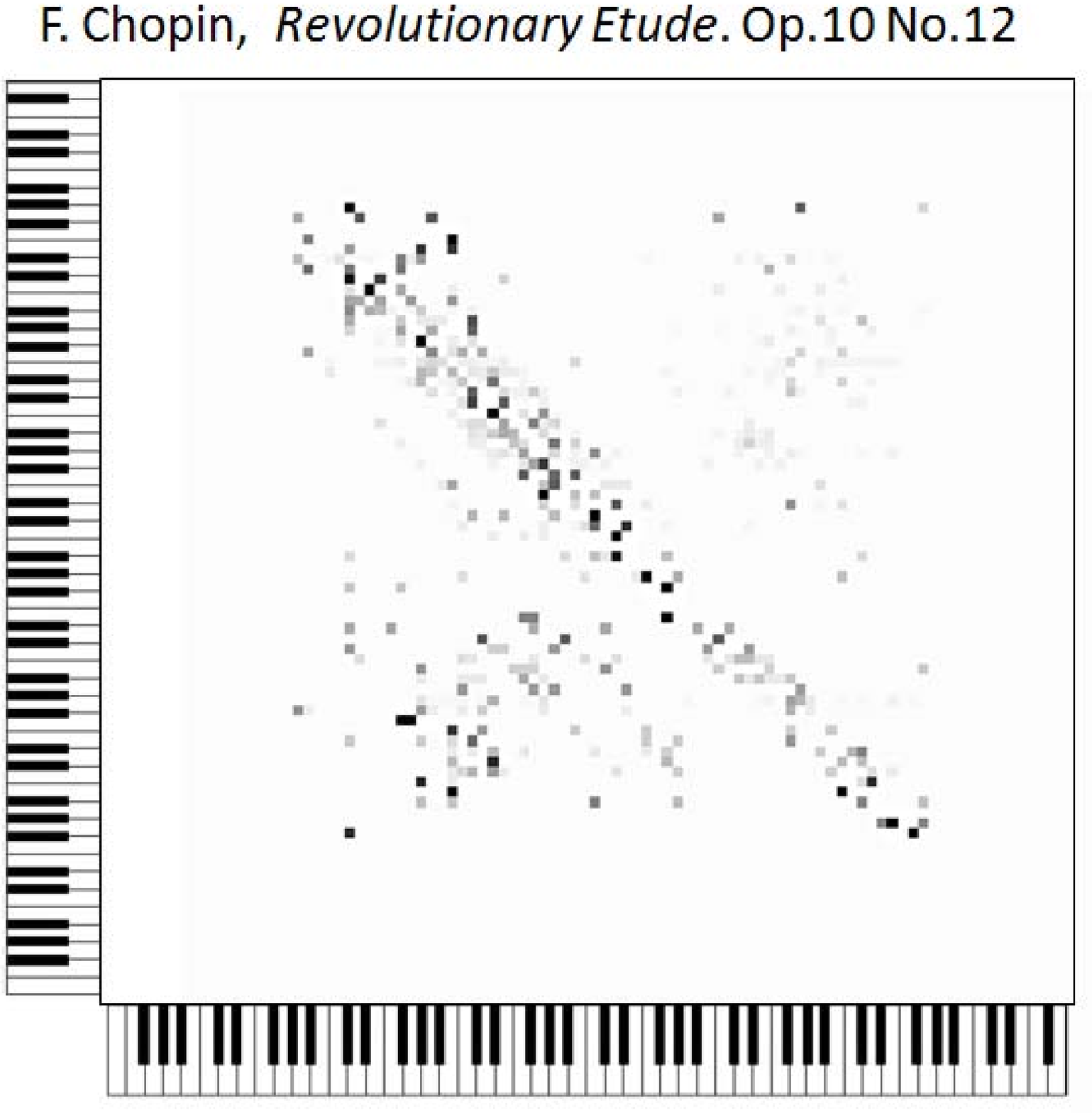, width=5.5cm, height =5.5cm}& 
 \epsfig{file=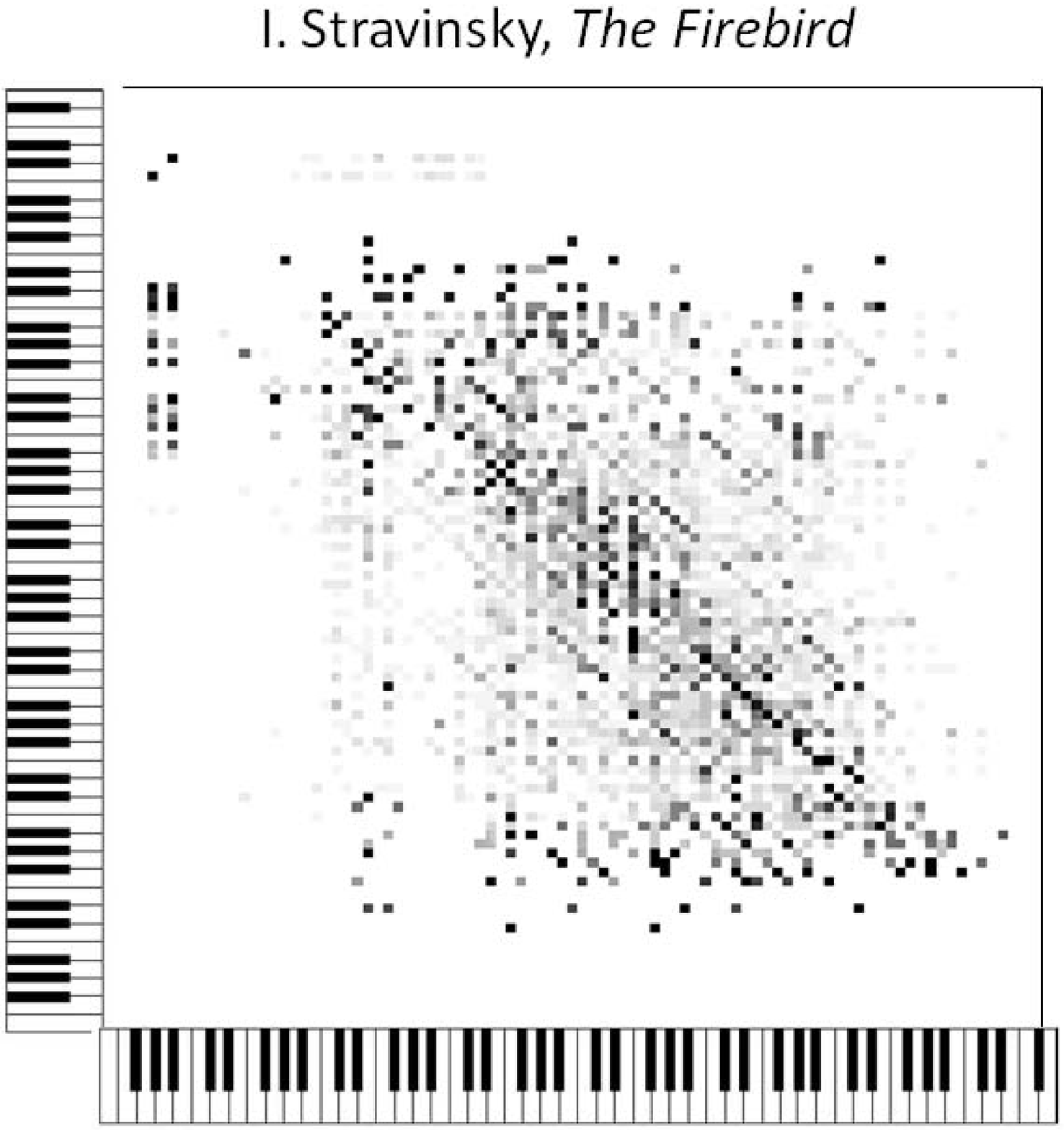, width=5.5cm, height =5.5cm} 
\end{tabular}
\end{center}
\caption{ Transition matrices for the MDG based on  
 the F.Chopin "Revolutionary Etude" (Op.10, No 12) (left)
and the I. Stravinsky "The Fire-bird" suite (right).
 \label{Fig_music_03}}
\end{figure}

We can consider a note as 
an elementary event 
providing 
a natural discretization 
of musical phenomena
that facilitate 
their performance and analysis.
Namely, 
given the entire keyboard $\mathcal{K}$
of 128 notes (standard for the MIDI representations of music)
corresponding to a pitch range of 10.5 octaves,
each divided into 12 semitones,
we regard a note  
as a discrete {\it random variable} $X$ 
that maps the musical event
 to a value of a $R$-set of pitches 
$\mathcal{P}=\{x_1, \ldots, x_R\}\subseteq \mathcal{K}.$
In the musical dice game,
a piece is generated by patching 
notes $X_t$ taking values from 
the set of pitches $\mathcal{P}$
that { sound good together}
into a  temporal  sequence
 $\left\{X_t\right\}_{t\geq 1}$.
Herewith, two consecutive notes, 
in which the second pitch is 
 a {\it harmonic} of the first one
 are considered to be pleasing to the ear,
and therefore can be patched
to the sequence. 
Harmony is based on 
 consonance, a concept whose definition  
changes permanently in musical history.
Two or more notes may sound
consonant for various reasons 
such as luck of perceptual 
roughness, spectral similarity 
of the sequence to a harmonic series, 
 familiarity of 
the sound combination in 
contemporary musical contexts,
and eventually for a personal taste,
as there are consonant and dissonant harmonies,
both of which are pleasing to the ears of some and not others.
A detailed statistical analysis 
of subtle harmony
conveyed by melodic lines
in  tonal music 
certainly calls for 
the complicated stochastic models, 
in which successive notes 
in the sequence $\left\{X_t\right\}_{t\geq 1}$
are not chosen independently,
 but their probabilities depend on preceding notes. 
In the general case,
a set of $n$-note probabilities
$\Pr\left[ X_{t+1}=x\mid X_t=y,X_{t-1}=z,\ldots,X_{t-n}=\upsilon\right]$
might be required to 
insure the resemblance 
of the musical dice games
to the original compositions.
However,
it is rather difficult 
to decide {\it a priori}
upon the enough memory depth $n$
in the stochastic models 
required 
to compare reliably 
the pieces of tonal and atonal music 
created by different composers,
with various purposes, 
in different epochs, 
for diverse musical instruments
subjected to the dissimilar tuning techniques.  
Under such circumstances, 
it is mandatory
to identify 
some meaningful blocks 
of musical information 
and to detect the hierarchical 
tonality (basic for perception of harmony in 
Western music \cite{Dahlhaus:2007})
in a simplified statistical model, 
as the first step of statistical analysis. 
For this purpose,
in the present work, 
we neglect possible statistical influences
extending over than the only preceding note
and limit our analysis
 to the simplest 
 time -- homogeneous
  Markov chain,
\begin{equation}
\label{music_transition_matrix}
\Pr\left[ X_{t+1}=x\mid X_t=y,X_{t-1}=z,\ldots\right]\,=\,
\Pr\left[ X_{t+1}=x\mid X_t=y\right]\,=\,T_{yx},
\end{equation}
where the elements of 
the stochastic transition matrix
$T_{yx},$ 
$\sum_{x\in \mathcal{P}}T_{yx}=1,$
weights the 
chance of a pitch $x$
going directly 
to another pitch 
$y$ independently of time.
It is worth mentioning that 
the model (\ref{music_transition_matrix})
obviously 
does not impose a severe limitation on  
melodic variability, 
since there are many possible 
combinations of notes considered consonant, 
as sharing some harmonics 
and making a pleasant sound together.
The relations between notes 
in (\ref{music_transition_matrix})
are rather described in terms of 
 probabilities 
and expected numbers of random steps 
 than by physical time.
Thus  
the actual length $N$ of a composition 
is formally put $N\to\infty,$ or
as long as you keep rolling the dice.
The examples of transition matrices generated 
for the different musical compositions are shown
in Fig.~\ref{Fig_music_03}. They 
appear to be essentially not symmetric:
if $T_{xy}>0,$ for some $x,y,$
 it might be that $T_{yx}=0.$
A musical composition can be represented 
by a weighted directed graph, in which 
vertices are associated with pitches 
and directed edges connecting them 
are weighted accordingly to the 
probabilities of the immediate transitions 
between those pitches.
  Markov's chains
determining random walks 
on such  graphs are not ergodic: 
 it may
  be impossible to go
 from every
 note  
to every other note 
following the score of the musical piece.

In music theory \cite{Thomson:1999},   
the hierarchical pitch relationships 
are introduced  based on a {\it tonic} key,
a pitch which is the lowest degree of a scale
and
that all other notes 
in a musical composition 
gravitate toward.
A successful tonal piece of music 
 gives a listener 
a feeling that 
a particular (tonic) chord 
 is the most stable and final.
The regular  method to establish
a tonic 
through a cadence, 
a succession of several 
chords which
 ends a musical section 
giving a feeling 
of closure,  
may be difficult to apply 
without listening to the piece.

While in a MDG, 
the intuitive vision of musicians
describing the tonic triad  
as  the "center of gravity" 
to which other chords are to lead
acquiers a quantitative expression.
Every pitch in a musical piece 
is characterized 
with respect to the entire structure of the Markov chain
by its level of accessibility 
estimated by the first passage time to it
\cite{Volchenkov:2011}
that is the expected length 
of the shortest path of a random walk
toward the pitch from any other pitch
randomly chosen over the musical score.
Analyzing the first passage times 
 in scores of tonal musical compositions,
we have found that they 
can help in resolving 
 tonality of a piece, 
as they precisely render
the hierarchical relationships between 
pitches. 
It is interesting to note that 
from the physical point of view
the first passage time can be naturally interpreted as
a potential,
as being 
equal to the diagonal elements of the generalized 
inverse of the Laplace operator. 
For example, 
the first passage time
to a node  precisely equals to 
the electric potential of the node, in an 
electric resistance network \cite{Volchenkov:2011}. 
Thus, in the framework of 
musical dice games, 
the role of a note 
in a tonal scale can be understood as its potential. 

\begin{center}
\begin{figure}[ht!]
 \noindent
\begin{tabular}{lr}
\epsfig{file=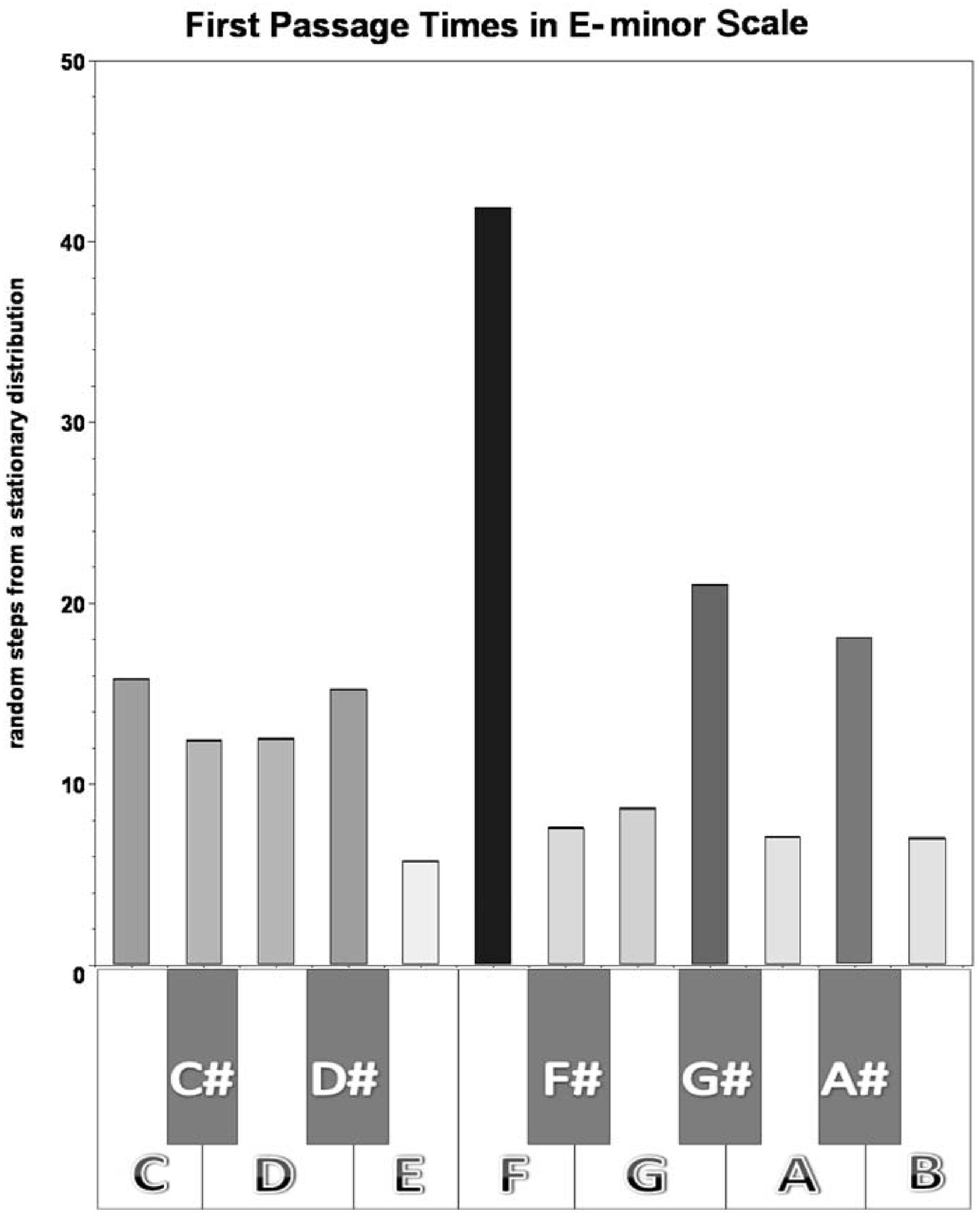, width=5.5cm, height =5.5cm}& 
 \epsfig{file=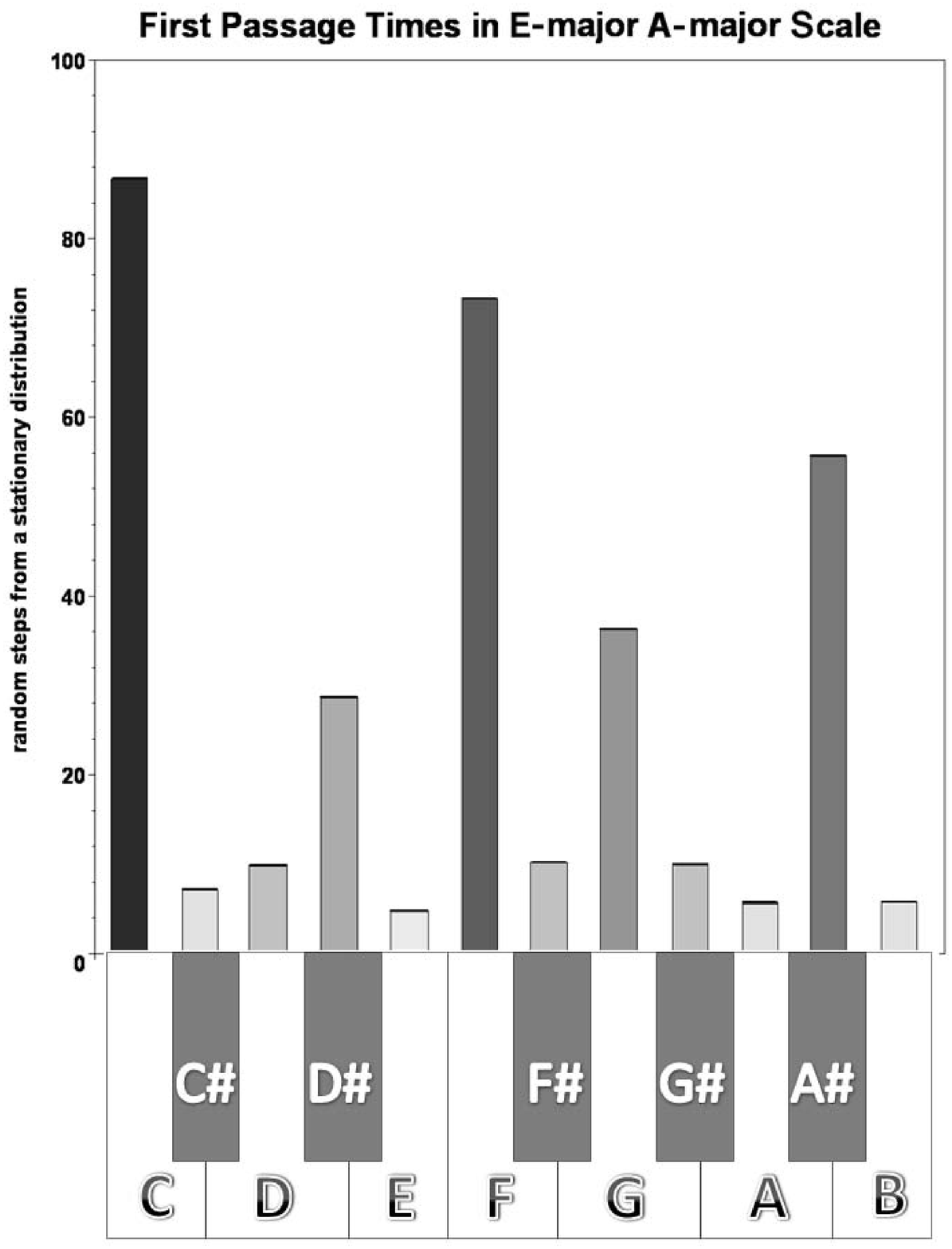, width=5.5cm, height =5.5cm}
\end{tabular}
\caption{ The histograms show the first passage times to the notes for the MDG over a part of Duet I of J.S. Bach (BWV 802) written in E minor  (left) and over a part of  the Cello Sonata No.3, Op.69 of  L.V. Beethoven written in E major, A  major (right) mapped into a single octave. Bars are shaded with the intensity of gray scale $[0-100\%]$,  in proportion to the magnitude of the first passage time. Therefore, the basic pitches of a tonal scale  are rendered with light gray color, as being characterized by short first passage times,  and the tonic key by the smallest magnitude of all. \label{fig6m}
}
\end{figure}
\end{center}

The majority of tonal music assumes that notes spaced 
over several octaves are perceived the same way 
as if they were played in one octave \cite{Burns:1999}. 

Using this assumption of 
octave equivalency,
we can 
 chromatically transpose
each musical piece into a single octave
getting  
the $12\times 12$ transition matrices,
uniformly for all musical pieces, independently 
of the actual number of pitches used in the composition.
Given a stochastic matrix ${\bf T}$
describing transitions 
between notes within a single octave $\mathcal{O}$, 
the first passage time to the note $i\in\mathcal{O}$
is computed in \cite{Volchenkov:2011}. 

We have shown in Fig.~\ref{fig6m} the two examples of 
the arrangements of first passage times 
to notes
in one octave, for the  
 E minor scale (left) and 
E major, A major scales (right).
The 
basic pitches for the 
E minor scale 
are 
\verb"E", \verb"F#", \verb"G", \verb"A",
\verb"B", \verb"C", and \verb"D".
The E major scale is based on \verb"E", \verb"F#",
\verb"G#", \verb"A", \verb"B",
\verb"C#",  and \verb"D#". 
Finally, the A major scale consists of
\verb"A", \verb"B", \verb"C#", \verb"D",
\verb"E", \verb"F#", and  \verb"G#".
The
values of first passage times 
 are strictly 
ordered in accordance to their role in 
the tone scale of the musical composition. 
Herewith, 
the tonic key is characterized 
by the shortest first passage time 
(usually ranged from 5 to 7 random steps),
and the values of first passage times 
to other notes collected in ascending 
order reveal the entire hierarchy of their 
relationships in the musical scale.

It is intuitive that
the time of recurrence  to a note
estimated by $\pi_i^{-1}$ 
is related positively to the first passage time
to it, $\mathcal{F}_i$:
the faster a random walk over the score
 hits the pitch for the first time,
the more often it might be expected to
occur again.
The time of recurrence  
equals the first passage time
 in a salient recurring
succession of notes (a motif) - 
the pattern of three short notes followed by one long
that opens the Fifth Symphony of L.V. Beethoven 
and reappears throughout the work is a classic example.

\begin{figure}[ht!]
 \noindent
\begin{center}
\epsfig{file=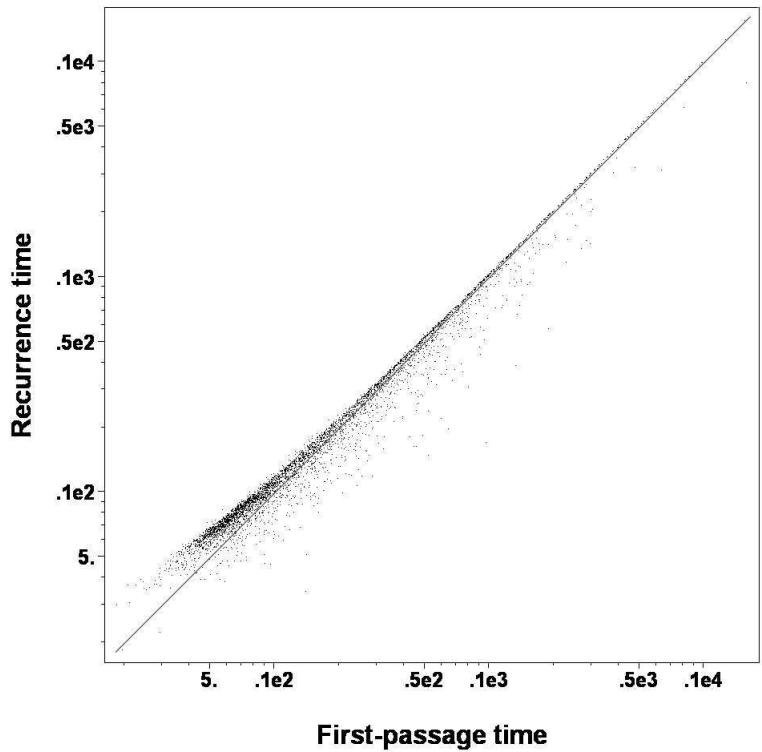,  angle= 0,width =10cm, height =10cm}
  \end{center}
\caption{
The log-log scatter plot contains $12\times 804$ points
representing the recurrence time vs. the first passage time to
the $12$ notes of one octave, over the MDG based on  
$804$ compositions of 
$29$ composers. The straight line is given for a reference
 indicating the horizon
 of intelligibility (when equality of recurrence times
 and first passage times  is achieved); 
 departures from linearity signify departures from
intelligibility.  \label{Fig_music_05x}}
 \end{figure}

The log-log scatter plot shown in Fig.~\ref{Fig_music_05x}
 represents the relation between the recurrence time and
 the first passage time to
the $12$ notes of one octave in all MDG over musical 
compositions we studied.
The straight line indicates equality of recurrence times
 and first passage times. The data 
 provide convincing evidence for the systematic 
departure of 
recurrence times from first passage times
for those pitches characterized by 
the relatively short first passage times (recurrence times 
to them are typically longer than first passage times).
The excess of recurrence times
over first passage times 
quantifies musical development 
encompassing 
distinct musical figures
that are subsequently altered
and sequenced throughout a piece of music.
It is not a surprise that 
such a musical development
is essentially visible 
in the range of  
the relatively short
first passage times, 
as they  
play 
an important role
 in the tonal scale structure 
of a piece guaranteeing its unity.

\section{Can we see first-passage times? \label{sec:cities}}
\noindent

A belief in the influence of the built environment on humans was common in architectural and urban thinking for centuries. Cities generate more interactions with more people than rural areas because they are central places of trade that benefit those who live there. People moved to cities because they intuitively perceived the advantages of urban life. City residence brought freedom from customary rural obligations to lord, community, or state and converted a compact space pattern into a pattern of relationships
by constraining mutual proximity between people.
Spatial organization
of a place
has an extremely important effect
on the way
people move through spaces
and meet other people by chance, \cite{Hillier:1984, Blanchard:2009}.
Compact neighborhoods can foster casual
social interactions among neighbors,
while creating barriers to interaction
 with people outside a neighborhood.
The phenomenon of clustering of minorities, especially that of newly
 arrived immigrants, is well documented \cite{Wirth}.
Clustering is considering to be beneficial for mutual support and
for the sustenance of cultural and religious activities. At the
same time, clustering and the subsequent physical segregation of
minority groups would cause their economic marginalization.
The spatial analysis of the immigrant
quarters \cite{Vaughan01}  and the
study of London's changes over 100 years \cite{Vaughan02}
shows that they were significantly more
segregated from the neighboring areas, in particular, the number
of street turning away from the quarters to the city centers were
found to be less than in the other inner-city areas being
usually socially barricaded by railways, canals and industries.
It has been suggested \cite{Language} that space structure and its
impact on  movement are critical to the link between the built
environment and its social functioning. Spatial structures creating a local situation
in which there is no relation between movements inside the spatial
pattern and outside it and the lack of natural space occupancy
become associated with the social misuse of the structurally
abandoned spaces.

In traditional
 urban researches, the dynamics of an urban pattern come
 from the landmasses, the physical aggregates of buildings
 delivering place for people and their activity.
The relationships between certain components of the urban texture
  are often measured along streets and routes considered as edges
  of a planar graph, while the traffic end points and street
  junctions are treated as nodes. Such a   primary graph
  representation of urban networks is grounded on relations
  between junctions through the segments of streets. The usual
   city map based on Euclidean geometry can be considered as an
   example of primary city graphs.
In space syntax theory (see \cite{Hillier:1984}),
 built environments are treated as systems
of spaces of vision subjected to a configuration analysis.
Being
irrelevant to the physical distances, spatial graphs
 representing the
urban environments are
removed from the physical space.
It has been demonstrated in multiple experiments
that spatial perception
shapes peoples understanding of how
a place is organized and eventually  determines the pattern of local
 movement, \cite{Hillier:1984}.
The decomposition of
urban spatial networks
into the complete sets
of intersecting open spaces
can be based on a number of different principles.
In  \cite{Jiang:2004},
while identifying a street over a plurality of routes
on a city map, the  named-street approach has been used, in
which two different arcs of the primary city network were
assigned to the same identification number (ID) provided they share the same
street name.

Being interested in the interpretation of urban space by pedestrians, 
in the present paper,
 we take a "named-streets"-oriented point of view
on the decomposition of
urban spatial networks
into the complete sets
of intersecting open spaces
 following our previous works \cite{Volchenkov:2007a,Volchenkov:2007b}.
 Being interested in the statistics of random walks defined on spatial
networks of urban patterns, we assign an individual
street ID code to each continuous segment of a street. The spatial
 graph of urban environment is then constructed by
mapping all edges (segments of streets) of the city map
shared the same street ID into nodes
 and all intersections among each pair of edges of the primary graph
into the edges of the secondary graph connecting the corresponding nodes.
We assign an individual
ID code to each continuous segment of a street or a channel. The spatial
 graph of urban environment is then constructed by
mapping all edges (segments of streets) of the city map
shared the same street ID into nodes
 and all intersections among each pair of edges of the primary graph
into the edges of the secondary graph connecting the corresponding nodes.

\begin{figure}[ht]
 \noindent
\epsfig{file=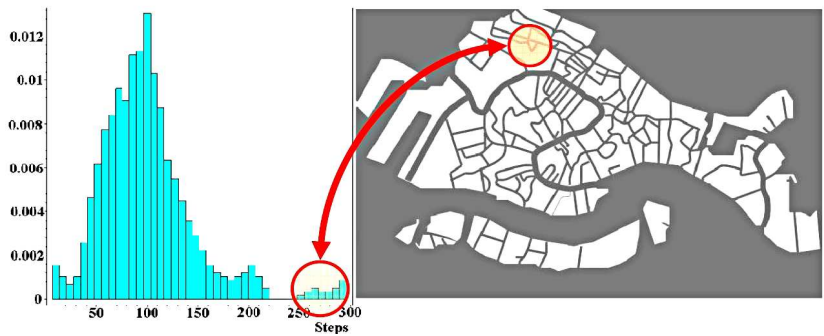,  angle= 0,width =11cm, height =4.5cm}
\caption{  The Venetian Ghetto jumped out
as by far the most isolated, despite being apparently well connected to the rest of the city. }
 \label{Fig3_urban}
\end{figure}

We have analyzed the first-passage times to individual canals in the
spatial graph of the canal network in Venice.
The distribution of numbers of canals
over the range of the first--passage time values is represented
by a histogram shown in Fig.~\ref{Fig3_urban} (left).
 The height of each bar in the histogram
is proportional to the number of canals in the
 canal network of Venice for which the first--passage
times fall into the disjoint intervals (known as bins).
Not surprisingly,
the Grand Canal, the giant Giudecca Canal
 and the Venetian lagoon are the most connected.
In contrast,  the Venetian Ghetto (see Fig.~\ref{Fig3_urban} (right)) -- jumped out
as by far the most isolated, despite being apparently well connected to the rest of the city --
 on average, it took 300 random steps to reach, far more than the average of 100 steps for other places in Venice.

The Ghetto was created in March 1516 to separate Jews from the Christian majority of Venice. It persisted until
1797, when Napoleon conquered the city and demolished the Ghetto's gates.
Now it is abandoned.

The notion of
isolation
 acquires
the statistical interpretation by means of random walks. The
first-passage times in the city vary strongly from  location to
location. Those places characterized by the shortest first-passage
times are easy to reach while very many random steps would be
required in order to get into a statistically isolated site.

Being a global characteristic
of a node in the graph,
the first-passage time
assigns  absolute scores
to all nodes
 based on the probability
 of paths they provide
 for random walkers.
The first-passage time
can therefore be considered
as a natural
measure of
statistical
isolation  of
 the node within the graph, \cite{Blanchard:2009}.
 \begin{figure}[ht!]
 \noindent
\epsfig{file=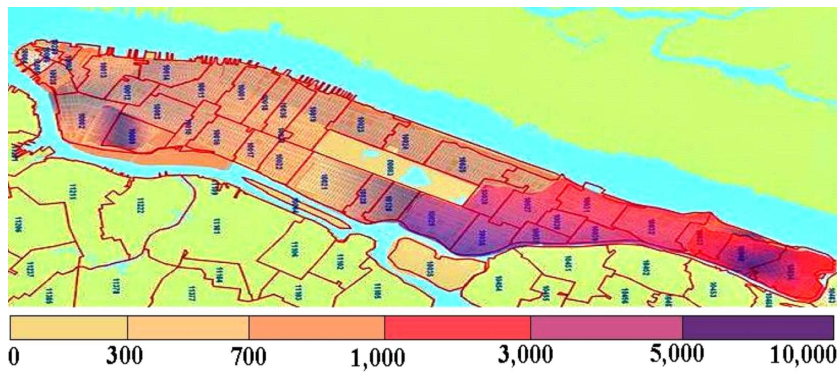, angle= 0,width =11cm, height =5cm}
\caption{ Isolation map of Manhattan. Isolation is measured
by  first-passage times to the places.  Darker color corresponds
to longer first-passage times.}
 \label{Fig2_Isolation}
\end{figure}

A  visual pattern
displayed on Fig.~\ref{Fig2_Isolation}
represents the pattern of structural
  isolation (quantified by the first-passage times)
 in Manhattan (darker color corresponds to longer first-passage times).
It is interesting to note that the  {spatial distribution of
isolation} in the urban pattern of Manhattan
(Fig.~\ref{Fig2_Isolation})
 shows a qualitative agreement with the map
of the  tax assessment  value of the land in Manhattan reported by
B. Rankin (2006) in the framework of the RADICAL CARTOGRAPHY
project being practically a negative image of that.

\begin{figure}[ht!]
 \noindent
\centering
\begin{tabular}{llll}
1). & \epsfig{file= 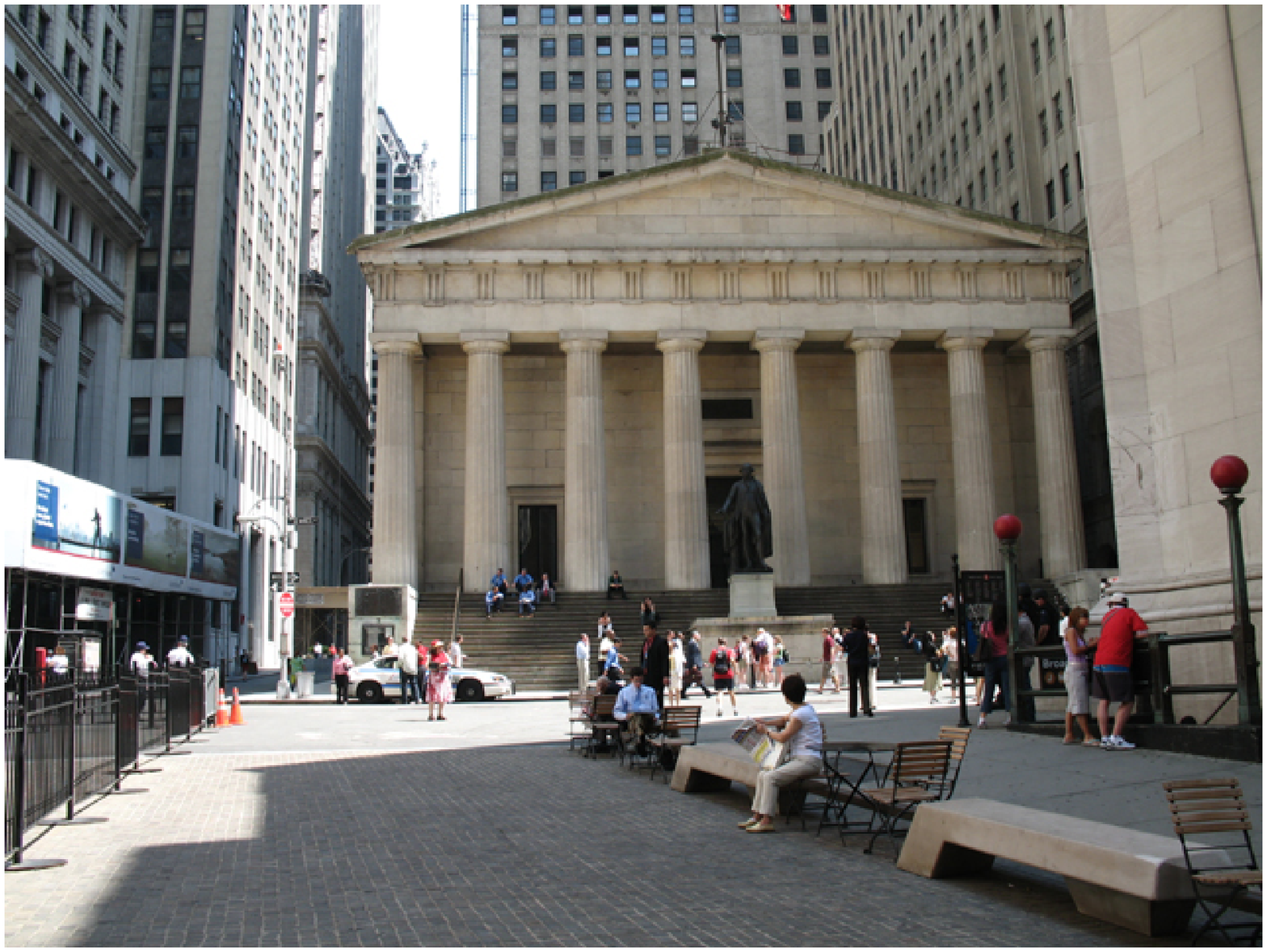, width=5.5cm, height =4cm}
& 2). &  \epsfig{file= 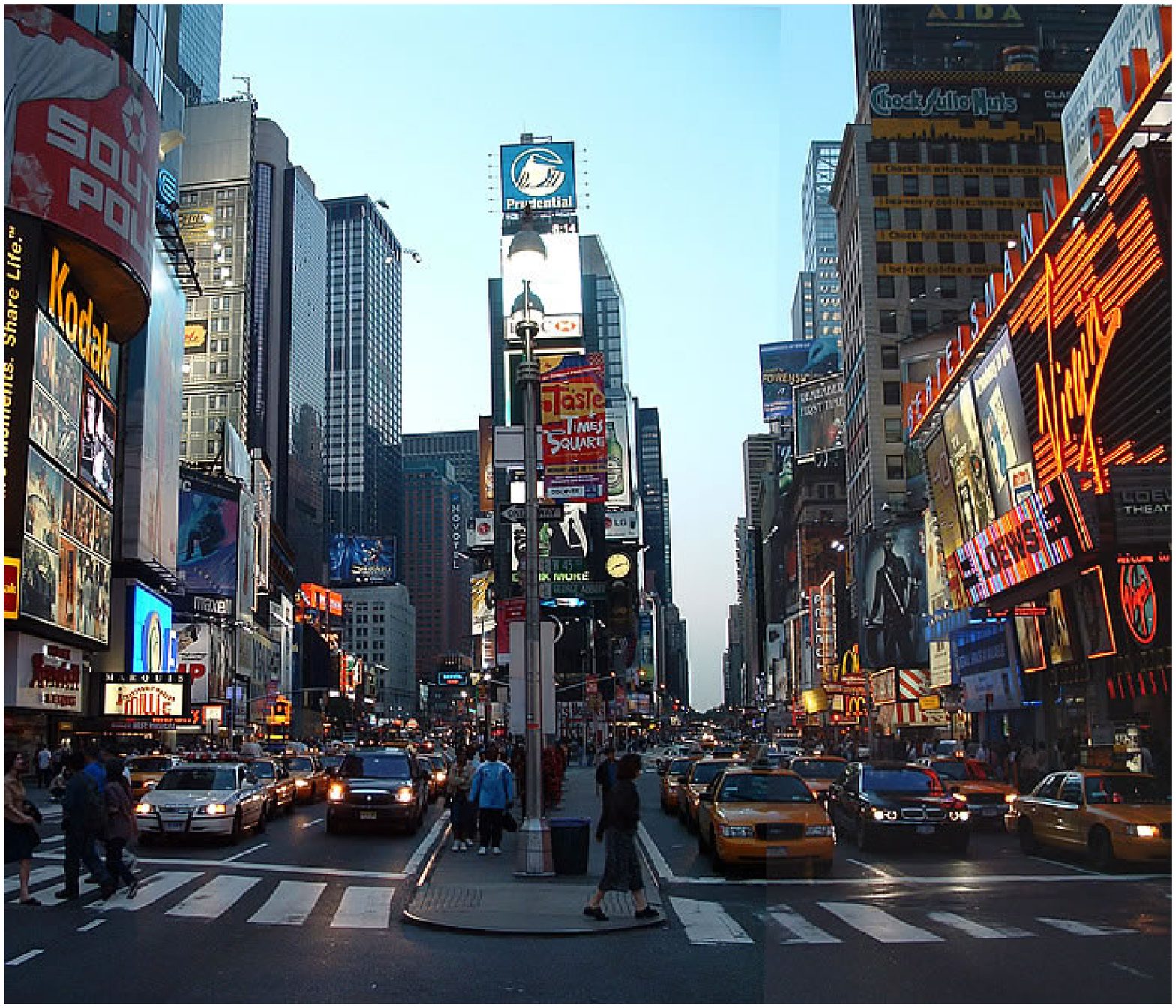, width=5.5cm, height =4cm} \\
3). & \epsfig{file= 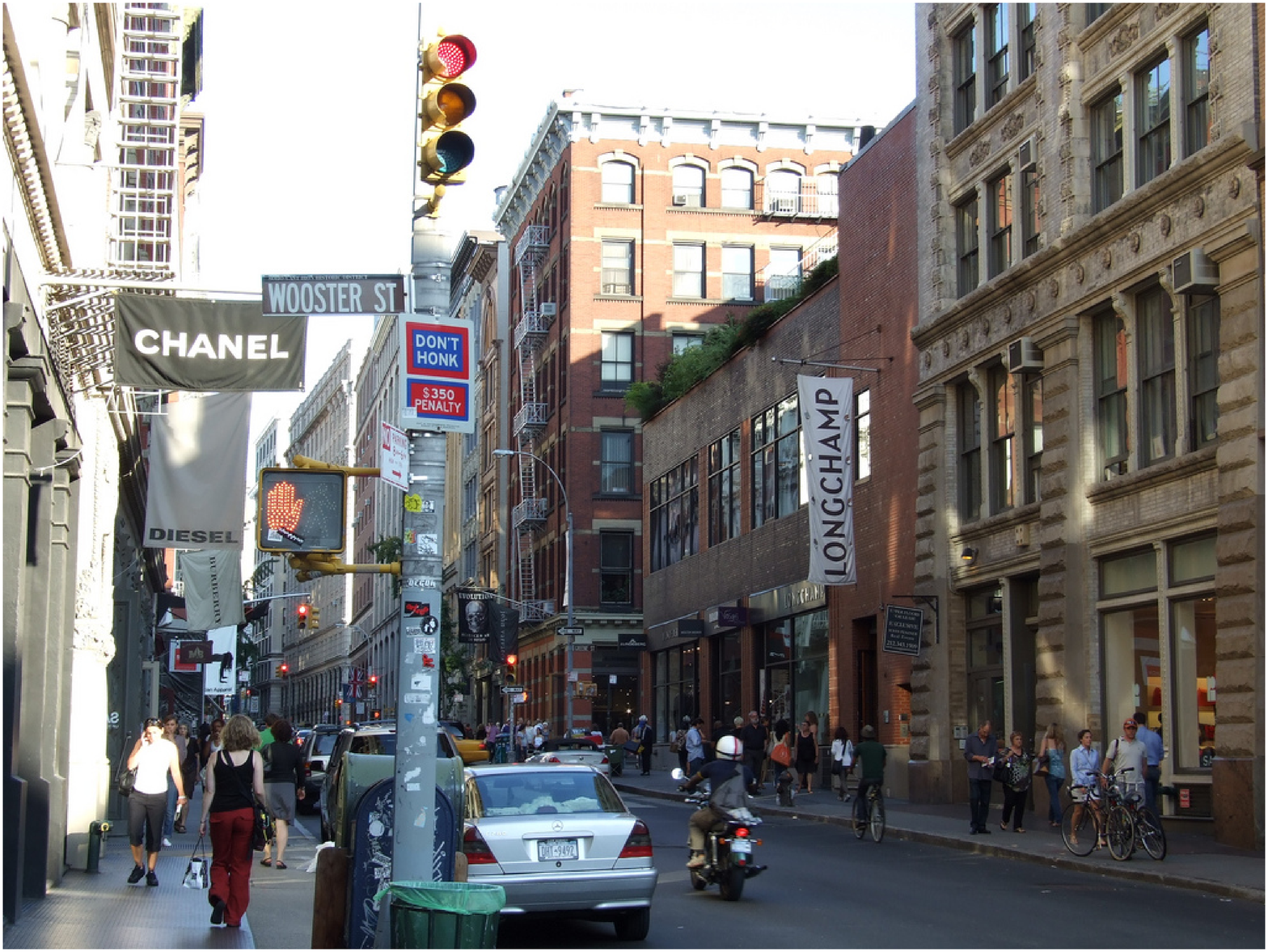, width= 5.5cm, height =4cm}
& 4). &  \epsfig{file= 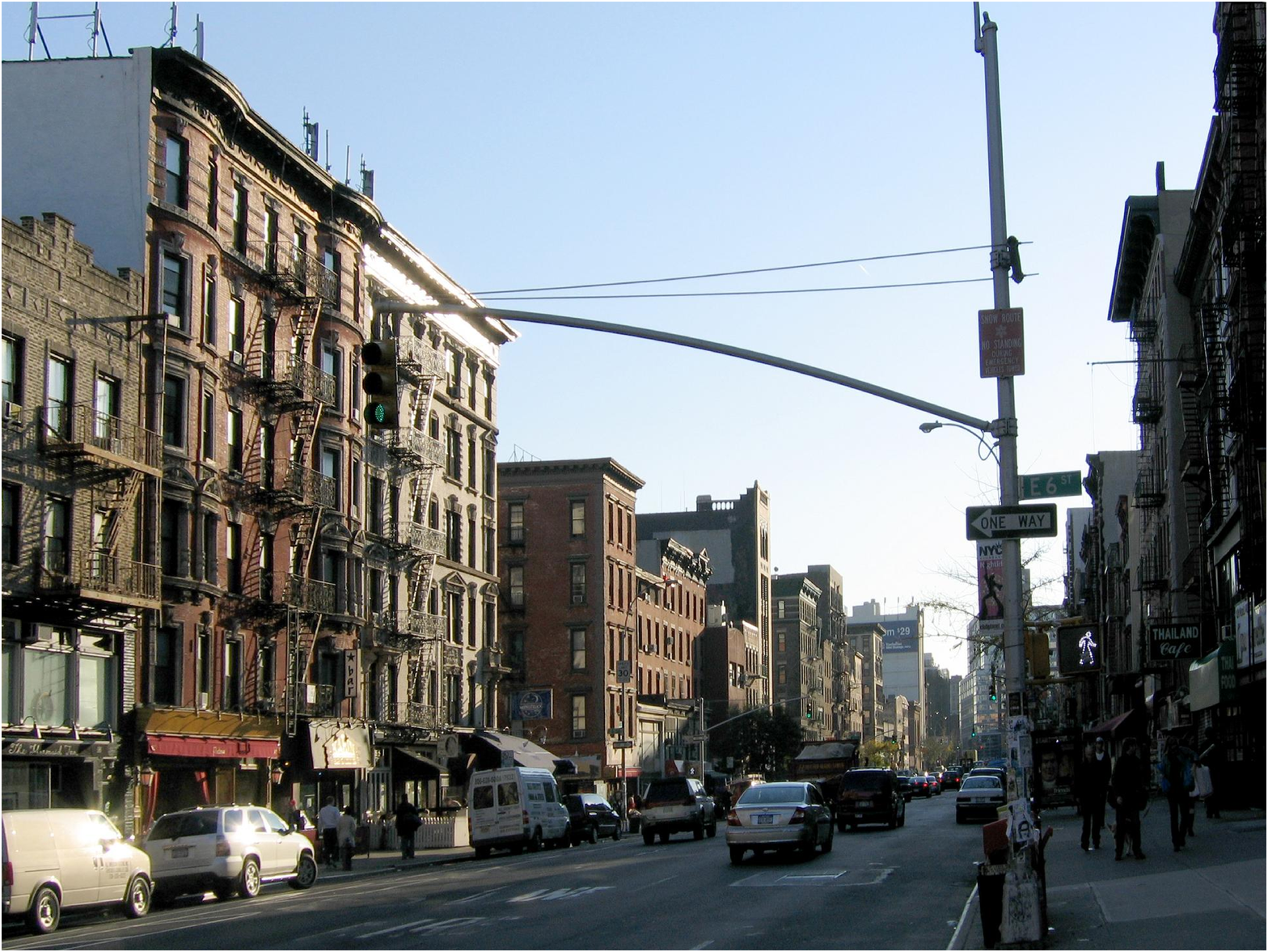, width=5.5cm, height =4cm} \\
5). & \epsfig{file= 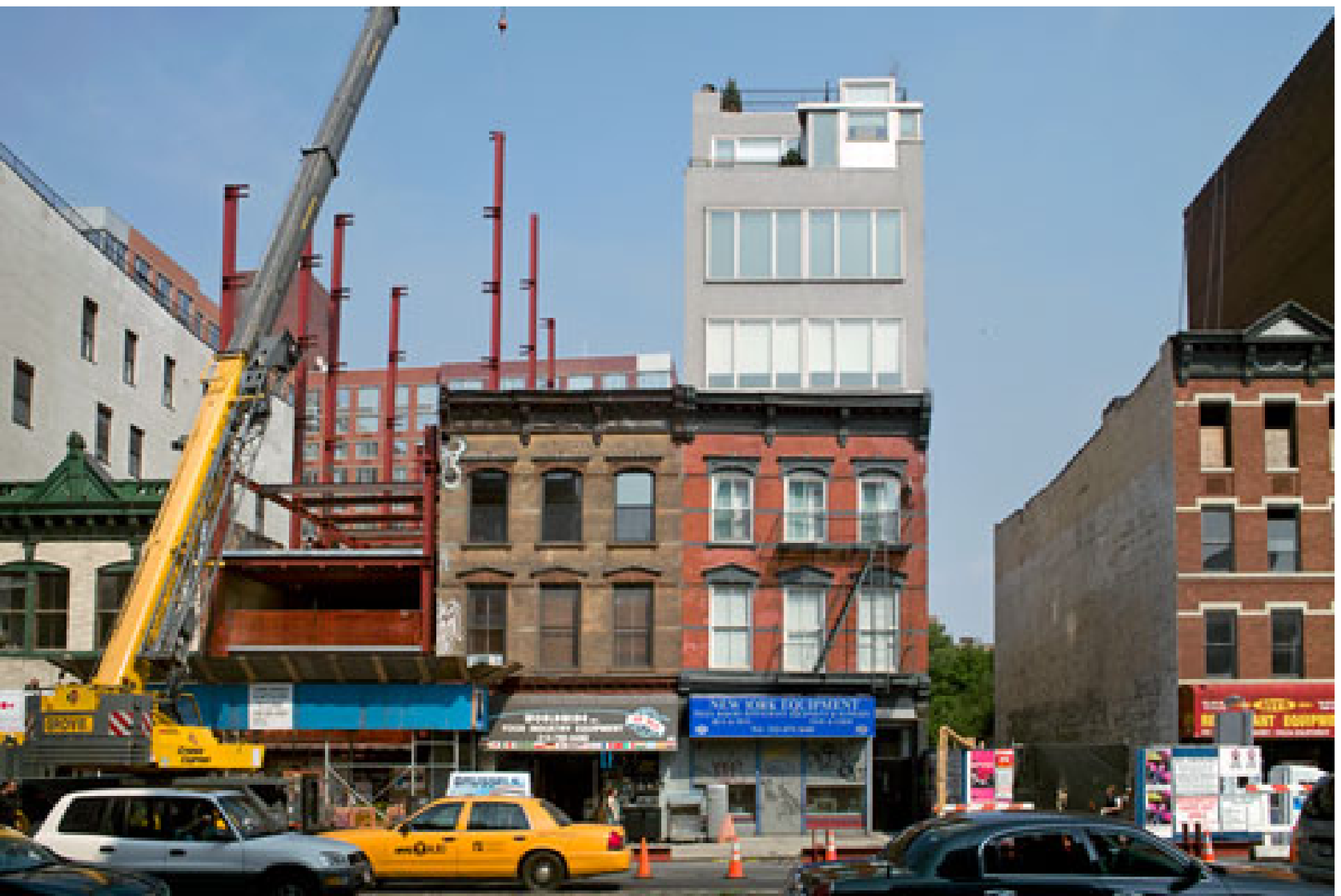, width= 5.5cm, height =4cm}
& 6). &  \epsfig{file= 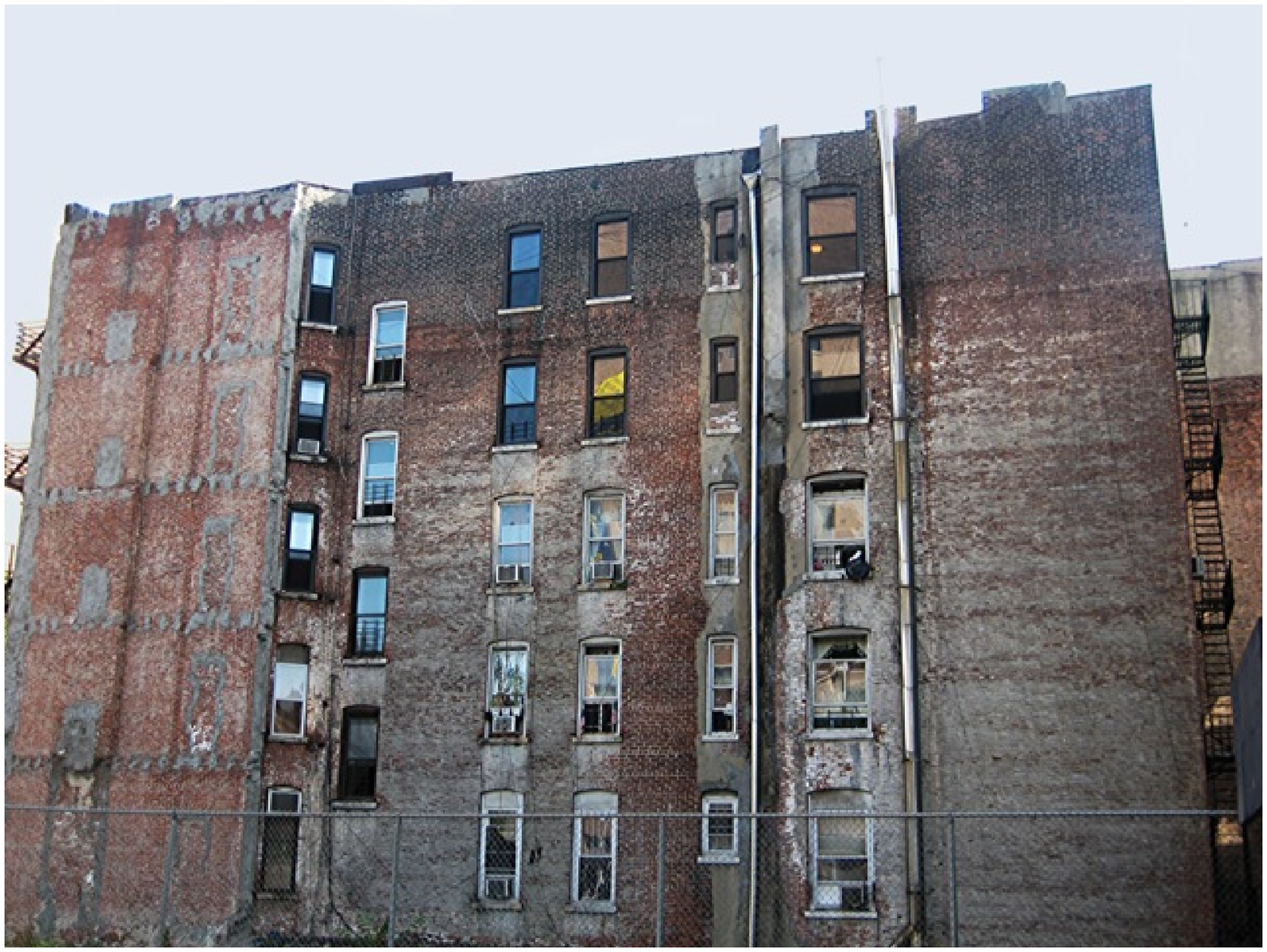, width=5.5cm, height =4cm} \\
\end{tabular}
\caption{ 
The first passage times in the borough of Manhattan, NYC: 
1). the Federal Hall National Memorial $\sim$ 10 steps; 
2). the Times square  $\sim$ 100 steps;
3). the SoHo neighborhood, in Lower Manhattan  $\sim$ 500 steps;
4). the East Village neighborhood, lying east of Greenwich Village, south of Gramercy and Stuyvesant Town  $\sim$ 1,000 steps;
5). the Bowery  neighborhood, in the southern portion of the New York City borough of Manhattan $\sim$ 5,000 steps;
6). the East Harlem (Spanish Harlem, El Barrio),  a section of Harlem located in the northeastern extremity of the borough of Manhattan $\sim$ 10,000 steps;
\label{Fig_Manhattan}}
\end{figure}

The  first-passage times enable us
to classify all places in the
 spatial graph of Manhattan
into four groups accordingly
to the first-passage times to them \cite{Blanchard:2009}.
The first group of locations
is characterized by the minimal first-passage times;
 they are probably reached for the first time
 from any other place
of the urban pattern  in just
 10 to 100 random navigational steps (the heart of the city), 
see Fig.~\ref{Fig_Manhattan}1 and Fig.~\ref{Fig_Manhattan}2.
These locations are identified as belonging to the downtown of
Manhattan (at the south and southwest tips of the island) -- the
Financial District and Midtown Manhattan. It is interesting to
note that these neighborhoods are roughly
coterminous with the
boundaries
 of the ancient New Amsterdam settlement founded in the
late 17${}^{\mathrm{th}}$ century.
Both districts comprise the offices and headquarters
of many of the city's major financial institutions
such as the New York Stock Exchange and the American Stock Exchange
(in the Financial District). Federal Hall National Memorial is also
encompassed in this area that had been
 anchored by the World Trade Center until the September 11, 2001
 terrorist attacks.
We might conclude that the group of locations characterized by the
best structural accessibility is the heart of the  {public process
in the city}.

The neighborhoods from the second
group (the  city core)
 comprise the locations
that can be reached for the first time
 in several hundreds
to roughly a thousand
 random navigational steps
from any other place of the urban pattern
(Fig.~\ref{Fig_Manhattan}3 and Fig.~\ref{Fig_Manhattan}4).
SoHo (to the south of Houston Street), Greenwich Village, Chelsea (Hell's Kitchen),
the Lower East Side, and the East Village
are among them --
they are commercial in nature and known for upscale shopping and
the "Bohemian" life-style of their dwellers
contributing into New York's art industry and nightlife.

The relatively isolated neighborhoods such
as Bowery (Fig.~\ref{Fig_Manhattan}5), some segments in Hamilton Heights and Hudson Heights,
Manhattanville (bordered on the south by Morningside Heights), TriBeCa
(Triangle Below Canal)
and some others can be associated to
 the third structural category
as being reached for the first time from 1,000 to 3,000 random
steps starting from a randomly chosen place in the spatial graph
of Manhattan. 
Interestingly, that
many locations belonging to the third structural
group comprises the
diverse and eclectic mix of different
social and religious groups.
Many famous  houses of worship
had been established there during the late
19${}^{\mathrm{th}}$ century --
St. Mary's Protestant Episcopal Church,
Church of the Annunciation,
St. Joseph's Roman Catholic Church,
and Old Broadway Synagogue in Manhattanville
are among them. The  neighborhood of  Bowery
in the southern portion of  Manhattan
had been most often associated with the poor and the homeless.
From the early 20${}^{\mathrm{th}}$ century, Bowery
became the center of the so called "b'hoy"
subculture of working-class
 young men
 frequenting the cruder nightlife.
Petty crime and prostitution followed in their wake, and
most respectable businesses, the middle-class,
 and entertainment had fled the area.
Nowadays, the dramatic decline has lowered crime rates in the
district to a level not seen since the early 1960s and continue to
fall. Although zero-tolerance policy targeting petty criminals is
being held up as a major reason for the crime combat success, no
clear explanation for the crime rate fall has been found.

The last structural category comprises
 the most isolated segments in the city
mainly allocated in the Spanish and East Harlems.
They are characterized by
the longest first-passage times from 5,000 to 
10,000  random steps (Fig.~\ref{Fig_Manhattan}6).
 Structural isolation is fostered
by  the
unfavorable confluence of many factors
 such as
the close proximity to Central Park
(an area of
340 hectares
removed from the otherwise regular street grid),
the
boundness by the strait of
 Harlem River
separating the Harlem and the Bronx,
and the remoteness from the main
bridges (the Triborough Bridge,
the Willis Avenue Bridge, and
the Queensboro Bridge) that
  connect the boroughs of Manhattan to
the urban arrays in
Long Island City and Astoria.
Many social problems associated with poverty
 from crime to drug addiction have plagued
the area for some time. The haphazard change of the racial
composition of the neighborhood occurred at the beginning of the
20${}^{\mathrm{th}}$ century together with the lack of adequate
urban infrastructure and services fomenting racial violence in
deprived communities and made the neighborhood unsafe -- Harlem
became a  {slum}. The neighborhood had suffered with unemployment,
poverty,  and crime for quite long time and even now, despite the
sweeping economic prosperity and redevelopment of many sections in
the district, the core of Harlem remains poor.

Recently, we have discussed in \cite{Volchenkov:2011}
that distributions of
 various social variables
(such as the mean household income and prison expenditures in
different zip code areas) may demonstrate the striking spatial
patterns which can be analyzed by means of random walks. In the
present work, we analyze the spatial distribution of the tax
assessment rate (TAR) in Manhattan.

The assessment tax relies upon a special enhancement made up of
the land or site value and differs from the market value
estimating a relative wealth of the place  within the city
commonly referred to as the 'unearned' increment of land use,
\cite{Bolton:1922}. The rate of appreciation in value of land is
affected by a variety of conditions, for example it may depend
upon other property in the same locality, will be due to a
legitimate demand for a site, and for occupancy and height of a
building upon it.

 \begin{figure}[ht!]
 \noindent
\begin{center}
\epsfig{file=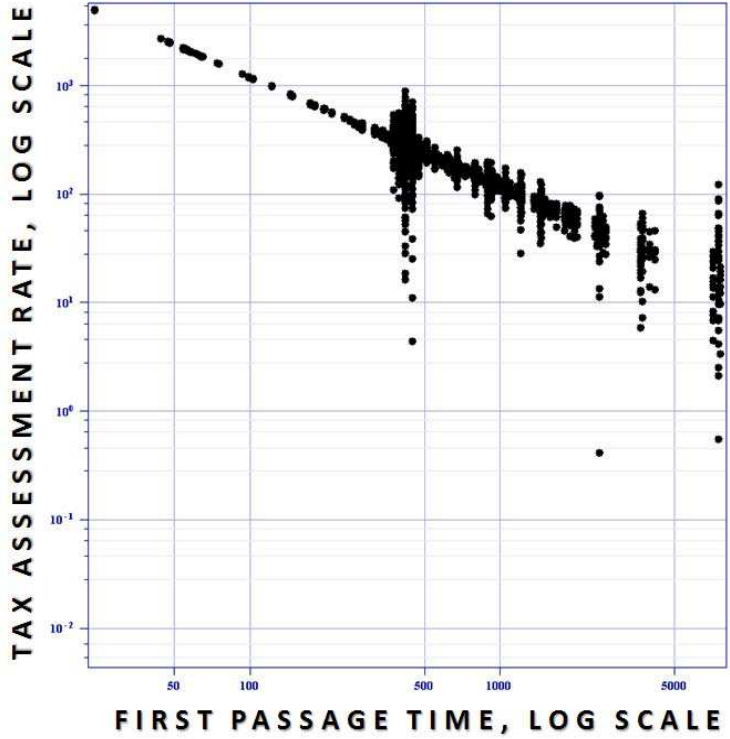, angle= 0,width =10cm, height =10cm}
  \end{center}
\caption{ Tax assessment rate (TAR)  of places in Manhattan
 (the vertical axes, in \$/fit${}^{2}$)
 is shown in the logarithmic scale
vs.
the first--passage times (FPT) to them (the horizontal axes).  }
 \label{Fig2_prices}
\end{figure}

The current tax assessment system enacted in 1981
in the city of New York
 classifies all real estate parcels into four classes subjected
to the different tax rates set by the legislature:
(i) primarily residential condominiums; (ii) other residential property;
(iii) real estate of utility corporations and special franchise properties;
(iv) all other properties, such as stores, warehouses, hotels, etc.
However, the scarcity
of physical space in the compact urban pattern on the island of Manhattan
will naturally set some increase of value on all desirably located
land as being a restricted commodity.
Furthermore, regulatory constraints on housing supply exerted on housing prices
by the state and the
city in the form of 'zoning taxes'
are responsible for
converting the property tax system in a complicated mess
of interlocking influences and for much of the high cost of housing
in Manhattan, \cite{Glaeser:2003}. 

Being intrigued with the
 likeness of
the tax
assessment map and the map of isolation in Manhattan,
 we have mapped the TAR figures publicly available
through the Office of the Surveyor at
the Manhattan Business Center onto the
 data on first-passage times to the corresponding
 places.
The resulting plot is shown in Fig.~\ref{Fig2_prices}, in the
logarithmic scale. The data presented in Fig.~\ref{Fig2_prices}
positively relates the geographic accessibility of places in
Manhattan
 with their 'unearned increments'
estimated by means of the increasing burden of taxation.
The inverse linear pattern dominating the data
is best fitted by the simple hyperbolic relation between
the tax assessment rate (TAR)
and the value of first--passage time (FPT),
$
\mathrm{TAR}\approx{c}/{\mathrm{FPT}},
$ in which $c\simeq 120,000\$\times\mathrm{Step}/\mathrm{fit}^2$ is a fitting
constant.

\section{First attaining times manifold. The Morse theory \label{sec:Morse}}
\noindent

In Sec.~\ref{sec:path_int}, we have shown that 
each node of a finite connected undirected graph,
 or a relational database 
can be described by 
a vector in the projective $(N-1)-$dimensional space, 
with the (squared) norm of the vector interpreted as 
the first-passage time to the node by 
the random walks,
starting from the {\it stationary distribution}. 
We have shown that the first-passage time 
can be calculated as the mean of all first hitting times,
$f_j=\sum_{i\geq 1}\pi_iH_{ij},$ 
with respect to the stationary distribution of  
random walks $\pi=\psi_1^2.$ For any given
 starting distribution $\phi_1^2$
that differs from the stationary one, 
we can also calculate the 
analogous quantity, 
\begin{equation}
\label{first_attaining_time}
\tilde{f}_j\left(\phi_1\right)\,\,=\,\,\sum_{i=1}^N\phi_{1,i}^2H_{i,j},
\end{equation}  
which we call the {\it first attaining} time to the node $j$ by the 
random walks starting at the distribution $\phi_{1}.$
In the present section, we demonstrate that 
first attaining times of a finite 
 undirected connected graph form a manifold 
(the {\it first attaining times} manifold)
in the $(N-1)-$dimensional probabilistic space, 
for which the first-passage times $f_j$ calculated 
with respect to the stationary distribution 
of random walks play the role of critical values. 
Below, we analyze 
 the topology of the 
first attaining times manifolds
by calculating their Euler characteristic 
and show that 
these manifold might have a fiber topological structure.

 Let us consider
the random walks starting 
from a distribution 
different from the stationary one,
\begin{equation}
\label{new_starting_distribution}
\phi_1\,\,=\,\,\left(1-\sum_{\kappa=2}^N\varepsilon_\kappa\right)\psi_1+
\sum_{\kappa=2}^N\varepsilon_\kappa\psi_\kappa,
\end{equation} 
in which $\varepsilon_\kappa>0$ is 
the magnitude of deviation 
of the vector $\phi_1$ from $\psi_1$
in the direction $\kappa$.
From the natural normalization condition 
for distributions,
$$
1\,\,=\,\,\sum_{i=1}^N\phi_{1,i}^2
\,\,=\,\,
\sum_{i=1}^N\left(
\left(1-\sum_{\kappa=2}^N\varepsilon_\kappa\right)\psi_{1,i}+
\sum_{\kappa=2}^N\varepsilon_\kappa\psi_{\kappa,i}
\right)^2,
$$
it follows that 
\begin{equation}
\label{normalization_2}
\left(1- \sum_{\kappa=2}^N\varepsilon_\kappa\right)^2\,\,=\,\,
1- \sum_{\kappa=2}^N\varepsilon^2_\nu, \quad |\varepsilon_\kappa|<1.
\end{equation}  
Then, the first attaining time 
for the node $j$ by the random walks 
starting  from the distribution $\phi_1$ 
is given by 
\begin{equation}
\label{fpt_new}
\begin{array}{ll}
\tilde{f}_{j}\,\, =\,\,\sum _{i=1}^{N}\phi _{1,i}^{2} H_{i,j} & =f_{j} \cdot \left(1-\sum _{\kappa =2}^{N}\varepsilon _{\kappa }  \right)^{2} \\ 
& +{\sum _{i=1}^{N}\left(2\left(1-\sum _{r=2}^{N}\varepsilon _{r}  \right)\sum _{\kappa =2}^{N}\varepsilon _{\kappa } \psi _{1,i} \psi _{\kappa ,i}  \sum _{s=2}^{N}\frac{1}{\lambda _{s} } \left(\frac{\psi _{s,j}^{2} }{\psi _{1,j}^{2} } -\frac{\psi _{s,j} }{\psi _{1,j} } \frac{\psi _{s,i} }{\psi _{1,i} } \right) \right) } \\
 & + \sum _{i=1}^{N}\left(\sum _{\kappa =2}^{N}\varepsilon _{\kappa } \psi _{\kappa ,i}  \right)^{2} \sum _{s=2}^{N}\frac{1}{\lambda  _{s} } \left(\frac{\psi _{s,j}^{2} }{\psi _{1,j}^{2} } -\frac{\psi _{s,j} }{\psi _{1,j} } \frac{\psi _{s,i} }{\psi _{1,i} } \right). 
\end{array}
\end{equation}
Taking into account the orthogonality condition for the 
eigenvectors $\psi_k$ and the relation (\ref{normalization_2}),
we further obtain,
\[\begin{array}{ll} 
\tilde{f}_{j}&{\,\,=\,\,f_{j} \cdot \left(1-\sum _{\kappa =2}^{N}\varepsilon _{\kappa }  \right)^{2} +f_{j} \cdot \sum _{i=1}^{N}\left(\sum _{\kappa =2}^{N}\varepsilon _{\kappa } \psi _{\kappa ,i}  \right)^{2}  -\sum _{i=1}^{N}\left(\sum _{\kappa =2}^{N}\varepsilon _{\kappa } \psi _{\kappa ,i}  \right)^{2} \sum _{s=2}^{N}\frac{1}{\lambda _{s} } \frac{\psi _{s,j} }{\psi _{1,j} } \frac{\psi _{s,i} }{\psi _{1,i} }   } \\
 &{\,\,=\,\,f_{j} \cdot \left(1-\sum _{\kappa =2}^{N}\varepsilon _{\kappa }  \right)^{2} +f_{j} \cdot \left(2\sum _{\kappa ,\nu =2}^{N}\left(1-\delta _{\kappa \nu } \right)\varepsilon _{\kappa } \varepsilon _{\nu } \underbrace{\sum _{i=1}^{N}\psi _{\kappa ,i} \psi _{\nu ,i}  }_{0}  +\sum _{\nu =2}^{N}\varepsilon _{\nu }^{2} \underbrace{\sum _{i=1}^{N}\psi _{s,i}^{2}  }_{1}  \right)-} \\
 &{\,\,\,\,-\sum _{i=1}^{N}\left(2\sum _{\kappa ,\nu =2}^{N}\left(1-\delta _{\kappa \nu } \right)\varepsilon _{\kappa } \varepsilon _{\nu } \psi _{\kappa ,i} \psi _{\nu ,i} \sum _{s=2}^{N}\frac{1}{\lambda  _{s} } \frac{\psi _{s,j} }{\psi _{1,j} } \frac{\psi _{s,i} }{\psi _{1,i} }   +\sum _{s=2}^{N}\varepsilon _{s}^{2} \psi _{s,i}^{2} \sum _{l=2}^{N}\frac{1}{\lambda  _{l} } \frac{\psi _{l,j} }{\psi _{1,j} } \frac{\psi _{l,i} }{\psi _{1,i} }   \right) } \\
& {\,\,=\,\,f_{j} \cdot \underbrace{\left(\left(1-\sum _{\kappa =2}^{N}\varepsilon _{\kappa }  \right)^{2} +\sum _{\nu =2}^{N}\varepsilon _{\nu }^{2}  \right)}_{1} -2\cdot \sum _{\kappa ,\nu =2}^{N}\varepsilon _{\kappa } \varepsilon _{\nu } \sum _{s=2}^{N}\frac{1}{\lambda   _{s} } \frac{\psi _{s,j} }{\psi _{1,j} } \sum _{i=1}^{N}\frac{\left(1-\delta _{\kappa \nu } \right)\psi _{\kappa ,i} \psi _{\nu ,i} \psi _{s,i} }{\psi _{1,i} }    } \\
& {{\rm \; \; \; }-\delta _{\kappa \nu } \sum _{\kappa =2}^{N}\varepsilon _{\kappa }^{2} \sum _{s=2}^{N}\frac{1}{\lambda  _{s} } \frac{\psi _{s,j} }{\psi _{1,j} } \sum _{i=1}^{N}\frac{\psi _{\kappa ,i}^{2} \psi _{s,i} }{\psi _{1,i} }.    } \end{array}\]     
Below, we introduce the following 
notations for the 
vectors in the $(N-1)-$dimensional projective space $P\mathbb{R}^{(N-1)}$, 
\begin{equation}
\label{Omega}
{\Omega}^{\kappa,\nu}\,\,=\,\,
\left\{
\frac 1{\sqrt{\lambda _s}}\sum_{i=1}^N 
\frac{\psi_{\kappa,i}\psi_{\nu,i}\psi_{s,i}}{\psi_{1,i}}
\right\}_{s=2}^N, \quad {\bf q}_i\,\,=\,\,
\left\{ 
\frac 1{\sqrt{\lambda _s}} \frac{\psi_{s,i}}{\psi_{1,i}}
\right\}_{s=2}^N  \,\,\in \,\,P\mathbb{R}^{(N-1)}, 
\end{equation}
and obtain the following expression for the 
first attaining times,
\begin{equation}
\label{new_first_passage}
\tilde{f}_{j}\,\, =\,\,{f}_{j} - \sum_{\kappa,\nu=2}^N\varepsilon_\kappa\varepsilon_\nu
\left({\bf q}_j,\Omega^{\kappa,\nu}\right)_{P\mathbb{R}^{(N-1)}}\,\,=\,\,
\left(
{\bf q}_j, {\bf q}_j -
\sum_{\kappa,\nu=2}^N\varepsilon_\kappa\varepsilon_\nu \Omega^{\kappa,\nu}
\right)_{P\mathbb{R}^{(N-1)}}
\end{equation}
where $(\bullet,\bullet)_{P\mathbb{R}^{(N-1)}}$ denotes the usual scalar product in $P\mathbb{R}^{(N-1)}.$
The second order derivative of the first attaining time
in the directions $\kappa,\nu$
(i.e., the {\it Hessian} function)
 is a symmetric matrix for
 any node $j$,
\begin{equation}
\label{Hessian}
{\mathcal H}_{\kappa\nu}(j)\,\,\equiv\,\,
\frac{\partial^2\tilde{f}_j}{\partial\varepsilon_\kappa\partial\varepsilon_\nu}
\,\,=\,\, -\left({\bf q}_j,\Omega^{\kappa,\nu}\right)_{P\mathbb{R}^{(N-1)}} \,\,=\,\, {\mathcal H}_{\nu\kappa}(j).
\end{equation}
Therefore,  
in a neighborhood of every node $j$, 
the first attaining time $\tilde{f}_j$ 
can be expressed as
\begin{equation}
\label{Euclidean_2}
\tilde{f}_j\,\,=\,\,
{f}_j + 
\left(
{\bf q}_j,\left(\sum_{\kappa,\nu=2}^N\varepsilon_\kappa\varepsilon_\nu
\frac{{\mathcal H}_{\kappa\nu}(j)}{\|{\bf q}_j\|_T}\right)\,{\bf q}_j
\right)_{P\mathbb{R}^{(N-1)}}.
\end{equation} 
where the first-passage time $f_j=\left({\bf q}_j,{\bf q}_j\right)_{P\mathbb{R}^{(N-1)}}$,
explicitly establishing the homeomorphism
to the Euclidean space of dimension $(N-1).$
We conclude that the first attaining times 
of all nodes of a finite connected undirected graph
form a manifold with respect to possible starting 
distributions of the random walks. 
It is important to mention that
the gradients of the 
first attaining times
to any node $j$ 
uniformly equal zero 
at the vicinity of the stationary distribution,
\begin{equation}
\label{gradients}
\nabla_\kappa\tilde{f}_j\equiv \left.
\frac{\partial \tilde{f}_j}{\partial{\varepsilon}_\kappa}\right|_{\varepsilon_\nu=0}\,\,=
\,\,\left.-\sum_{\nu=2}^N\varepsilon_{\nu}
\left({\bf q}_j,\Omega^{\kappa,\nu}\right)_{P\mathbb{R}^{(N-1)}}\right|_{\varepsilon_\nu=0} \,\,=\,\, 0.
\end{equation}
Therefore, each node $j$ is 
a critical point of the manifold
of first attaining times, and
the first passage times 
$f_j$ are the correspondent
 critical values.
Furthermore, 
following the ideas 
 of the Morse theory \cite{Matsumoto:2002},
we can perform the standard classification of the 
critical points, saying 
that a critical point $j$
is non-degenerate
if the Hessian matrix (\ref{Hessian}) is 
non-singular
at $j$ and 
introducing 
the {\it index} $\gamma_j$ of 
the non-degenerate critical point $j$ 
as the number of negative eigenvalues of 
(\ref{Hessian}) at $j$
(i.e., the dimension of the largest subspace 
of the tangent space to the manifold at $j$
 on which the Hessian is negative definite). 

\begin{figure}[ht!]
 \noindent
\centering
\begin{tabular}{cc}
\epsfig{file= 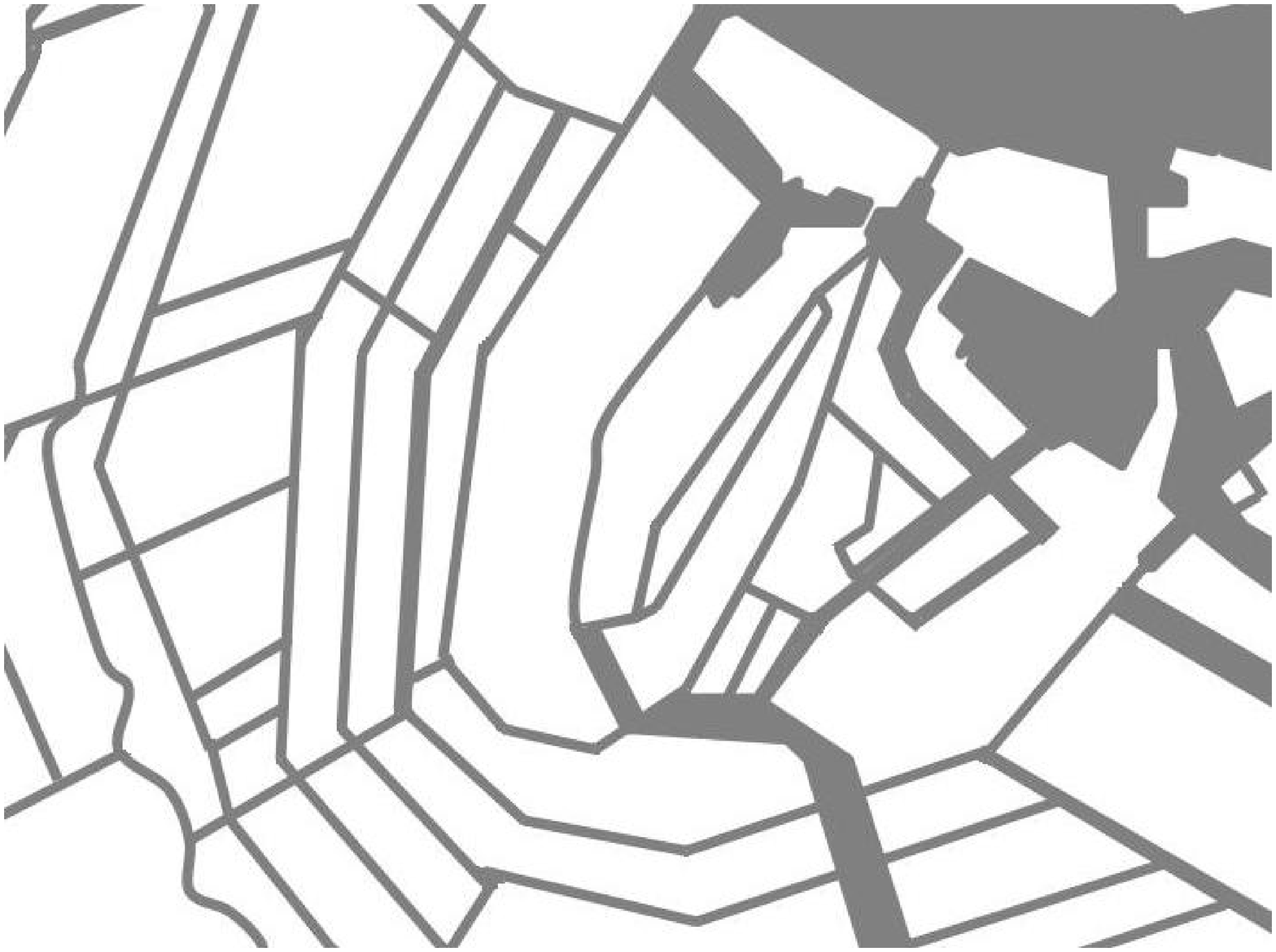, width=3.5cm, height =3.0cm} & 
\epsfig{file= 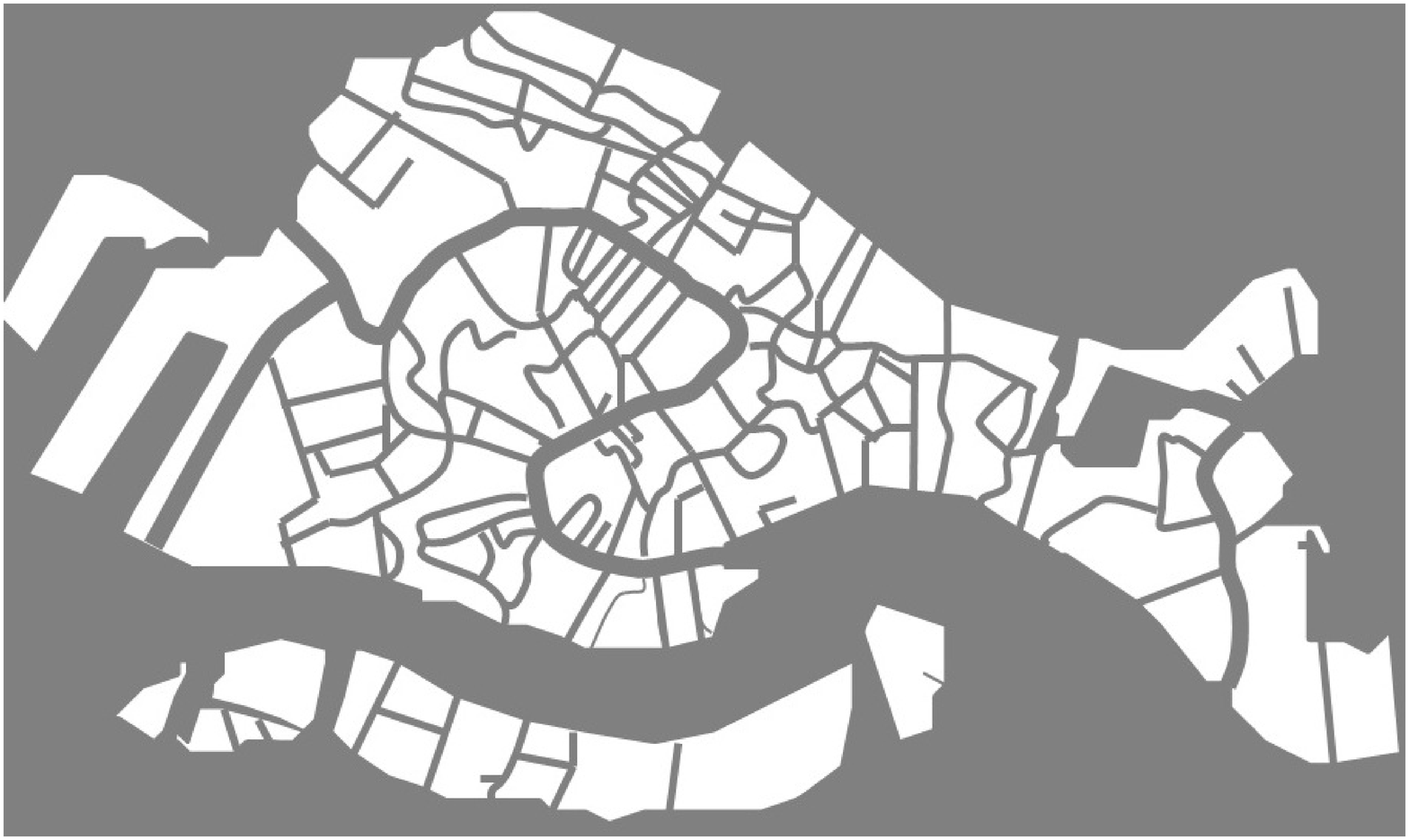, width=5cm, height =3cm} \\
$\chi=-5$   &   $\chi=4$
\end{tabular}
\caption{ The 
Euler characteristics
for the first attaining times manifolds
calculated for
the webs of city canals in
Amsterdam (57 canals) and
 Venice (96 canals). \label{Fig_Euler}}
\end{figure}

The Euler characteristic $\chi$ is 
an intrinsic property of a manifold
that describes its topological space's 
shape  regardless of the way it is bent.
It is known that 
the Euler characteristic 
can be calculated as 
the alternating sum of 
$C^{\gamma},$ 
the numbers of critical points
 of index $\gamma$ of the Hessian function,
\begin{equation}
\label{chi}
\chi\,\,=\,\, \sum_{\gamma\geq 0} (-1)^\gamma C^{\gamma}.
\end{equation}
In Fig.~\ref{Fig_Euler},
we have presented the 
Euler characteristics
for the first attaining times manifolds
calculated for
the webs of city canals in
Amsterdam (57 canals) and 
Venice (96 canals).
It is interesting to note that 
the functions of the canal networks
created in Venice and in Amsterdam 
were historically different. 
While the Venetian canals
mostly serve the function 
of transportation routes between the distinct districts of
the gradually growing naval capital of the Mediterranean region, 
the concentric web of Amsterdam gratchen had been built 
in order to defend the city.
The central diamond within a walled city was thought to be a good design for defense. Many Dutch cities have been structured this way: a central square surrounded by concentric canals. The city of Amsterdam is located on the banks of the rivers Amstel and Schinkel, and the bay IJ. It was founded in the late 12th Century as a small fishing village, but the concentric canals were largely built during the Dutch Golden Age, in the 17th Century.  The principal canals are three similar waterways, with their ends resting on the IJ, extending in the form of crescents nearly parallel to each other and to the outer canal. Each of these canals marks the line of the city walls at different periods. Lesser canals intersect the others along the radial direction, dividing the city into a number of islands. 
The resulting structure is highly symmetric,
 and the Euler characteristic of the corresponding 
first attaining times manifold is negative ($\chi=-5$). 
It is known that  the
negative Euler characteristics could either come 
from a pattern of symmetry in the hyperbolic surfaces, or 
from a manifold homeomorphic multiple tori.

In contrast to the highly symmetric and planned 
structure of the canal network in Amsterdam, 
the network of Venetian canals was developed 
gradually, in a long historical process 
of urban development
started from the 6th Century, at the  
oldest settlements in Venice that had appeared  in
Dorsoduro, along the Giudecca Canal. 
By the 11th Century, settlement had spread to the Grand Canal. The Giudecca island was composed of eight islets separated by
canals dredged in the 9th Century when the area was 
divided among the rebelling nobles.
The primary traditional divisions of Venice
into  sestieri (Cannaregio, San Polo, Dorsoduro,
 Santa Croce, San Marco and Castello, Giudecca) 
was developed from 
the early Middle Ages to the 20th Century.
The Euler characteristic of the 
first attaining times manifold
corresponding to the resulting network 
of Venetian canals
 is $\chi=4$. 
The large positive value of the Euler characteristic
can arise due to the well-known 
product property of Euler characteristics 
   $\chi(M \times N) = \chi(M) \cdot \chi(N)$,
for any product space $M \times N,$ or,
more generally, 
 from a fibration, 
when one topological space (called a fiber) is 
being "parameterized" by another topological space (called a base).

\section{Discussion and Conclusion \label{sec:discussion}}
\noindent

In the present paper, in order to geometrize and interpret the data,
we propose the use of a path integral distance, in which all possible paths between any two nodes in a graph model of the data are taken into account, although some paths are more preferable than others. We have pointed out that 
the process of data interpretation is always based on the implicit introduction 
of certain equivalence relations on the set of walks over the database.
It is clear that different equivalence relations might lead to the different concepts of taxonomic categories and generate the different 
concept for representing the data. Therefore, an interpretation process does not necessary reveal a "true meaning" of the data, but rather represents a self-consistent point of view on that. 

Given a probability measure on the set of all walks in the graph model
 of the database, every equivalence relation defines the discrete time random walks of a certain scale, in which all equivalent walks are equiprobable. 
We have shown that the random walks of different scales are dramatically dissimilar: while the nearest-neighbor random walks homogeneously cover the whole available space within the given environment, the high scale random walks
are obstacle repelling, as the centrality measure (quantifying the fraction of paths of the length $n>1$ traversing the node) is sensitive to defects and 
dramatically decreases in the vicinity of obstacles and boundaries. As the scale of the random walk increases, 
the level of transition randomness quantified by the 
entropy rate  shows almost the logarithmic growth, approaching 
its maximum for the maximum entropy random walks $T^{(\infty)}.$
At the same time, the behavior of the excess entropy ("complexity", 
"the past-future mutual information") is complementary, decreasing from 
its maximal value attained for the nearest-neighbor random walks
to the walks characterized 
by the minimal dependence on the past path attained for the maximum entropy random walks $T^{(\infty)}.$
In order to study the relation between the order of transitions and the structural correlations within an environment by the complexity -entropy relations in the three different environments of increasing structural 
complexity. We have shown that 
the relation between structure and order in the random transitions within an 
"empty room" is straightforward across the different scales: the structure provided merely by the walls, bounding the environment reinforces the order. As the number of obstacles within the environment increases, such the reinforcement becomes even more visible, especially for the large scale random walks. As the structural complexity of the environment further increases, the phenomenon of localization of random walkers inside the spaces formed by the filamentary sequences of the closely located obstacles comes into play at the short-range random walks $T^{(2)}.$

The random walks allow us to geometrize the data by introducing the new graph distance, in which all possible paths between any two nodes are taken into account. The idea of the path integral generalizes the action principle of classical mechanics  replacing a single trajectory of a particle with a sum over all possible trajectories to compute the probability amplitude (propagator).  The path integral calculation can be viewed as the left inverse of the Laplace operator describing the diffusion process of particles. The inverse of the 
Laplace operator describing the spread of particles in an environment  is not unique, however we can choose such the generalized inverse of it which share the same set of symmetries as the Laplace operator itself (the Drazin generalized inverse). It is important to mention that the calculation of the Green's function for the diffusion process give rise to a non-trivial geometric framework for description of graphs and relational databases. 
In particular, each node of the finite connected undirected graphs 
can be described by a vector in the projective $(N-1)-$dimensional space, with the squared norm of the vector interpreted as the first-passage time to the node by the random walks, starting from the stationary distribution. The image of the graph in such the probabilistic projective space constitutes a diffusion manifold of self-avoiding random walks in affine space, as 
we can subtract the graph nodes (by component wise subtracting of their images) to get vectors, or add vectors to a vertex to get another vertex, but we cannot add new vertices.
In our paper, we consider some examples of the use of the probabilistic geometry for the description of structural properties of mazes, musical compositions, and the various urban neighborhoods. We have found that the relation between the recurrence time and the first -passage time to the node can play the crucial role for the interpretation of its role within the structure. Namely, the isolated places for which the first-passage times are longer than the recurrence times to them might be interpreted as obscure, since the walker appears to be trapped within them. On the contrary, 
the nodes characterized with the first-passage times shorter than the recurrence time can play the important role, providing the guide to the environment 
and increasing its intelligibility. One fascinating example 
 is the tonality structure in the western musical compositions. 

Being interested in the interpretation of urban spatial patterns, we have analyzed the first-passage times in the canal network of Venice and the urban pattern on Manhattan. The first -passage times enable us to classify 
all places of movement in the urban spatial graph accordingly to their 
statistical isolations. It is remarkable that the obtained classification of places correspond to the spatial distribution of the tax assessment rate of sites that relies upon a special enhancement made up of the site value.

Finally, we consider the quantity similar to the first- passage times, which we have called the first attaining times, quantifying the first -passage time to the node by the random walk starting from a distribution that differs from the stationary one. We have shown that the set of the first attaining times form a manifold within the projective probabilistic space, for which the first-passage times plays the role of critical values. The Euler characteristics of the studied real-world graphs (canal networks in Venice and Amsterdam) show that these manifold might have a fiber topological structure.

\section*{Acknowledgement}
\noindent

Financial support by the project 
{\it MatheMACS}
 ("Mathematics of Multilevel Anticipatory Complex Systems"),
 grant agreement no. 318723,
funded by the EC Seventh Framework Programme
 FP7-ICT-2011-8
  is gratefully acknowledged.

\end{document}